\documentclass[useAMS,usenatbib]{mn2e}
\usepackage{graphicx}
\usepackage{multirow}
\usepackage{amsmath}

\title[Metal-poor stars in the Galactic bulge]{The EMBLA Survey - Metal-poor stars in the Galactic bulge\thanks{This paper includes data gathered with the 6.5 meter Magellan Telescopes located at Las Campanas Observatory, Chile.}}
\author[Louise M. Howes et al.]{Louise M. Howes,$^{1,2}$\thanks{E-mail: louise@astro.lu.se} Martin Asplund,$^{1}$ Stefan C. Keller,$^{1}$ Andrew R. Casey,$^{3}$ David Yong,$^{1}$ \newauthor 
Karin Lind,$^{4}$ Anna Frebel,$^{5}$ Austin Hays,$^{5}$ Alan Alves-Brito,$^{6}$ Michael S. Bessell,$^{1}$ \newauthor
Luca Casagrande,$^{1}$ Anna F. Marino,$^{1}$ David M. Nataf,$^{1}$ Christopher I. Owen,$^{1}$ \newauthor
Gary S. Da Costa,$^{1}$ Brian P. Schmidt,$^{1}$ Patrick Tisserand$^{1,7}$
\\
$^{1}$Research School of Astronomy and Astrophysics, Mt. Stromlo Observatory, Cotter Road, Weston, ACT 2611, Australia\\
$^{2}$Lund Observatory, Department of Astronomy and Theoretical Physics, Lund University, Box 43, SE-22100 Lund, Sweden\\
$^{3}$Institute of Astronomy, University of Cambridge, Madingley Road, Cambridge, CB3 0HA, United Kingdom\\
$^{4}$Department of Physics and Astronomy, Division of Astronomy and Space Physics, Uppsala University, Box 516, SE-751 20 Uppsala, Sweden\\
$^{5}$Massachusetts Institute of Technology, Kavli Institute for Astrophysics and Space Research, Cambridge, MA 02139, USA\\
$^{6}$Instituto de F\'isica, Universidade Federal do Rio Grande do Sul, 91501-970 Porto Alegre, RS, Brazil\\
$^{7}$Sorbonne Universit\'es, UPMC Univ Paris 6 et CNRS, UMR 7095, Institut d'Astrophysique de Paris, 98 bis bd Arago, 75014 Paris, France
}
\begin{document}

\date{}

\pagerange{\pageref{firstpage}--\pageref{lastpage}} \pubyear{2015}

\maketitle

\label{firstpage}

\begin{abstract}
Cosmological models predict the oldest stars in the Galaxy should be found closest to the centre of the potential well, in the bulge.  The EMBLA Survey successfully searched for these old, metal-poor stars by making use of the distinctive SkyMapper photometric filters to discover candidate metal-poor stars in the bulge. Their metal-poor nature was then confirmed using the AAOmega spectrograph on the AAT. Here we present an abundance analysis of 10 bulge stars with $-2.8$$<$[Fe/H]$<$$-1.7$ from MIKE/Magellan observations, in total determining the abundances of 22 elements.  Combining these results with our previous high-resolution data taken as part of the Gaia-ESO Survey, we have started to put together a picture of the chemical and kinematic nature of the most metal-poor stars in the bulge. The currently available kinematic data is consistent with the stars belonging to the bulge, although more accurate measurements are needed to constrain the stars' orbits. The chemistry of these bulge stars deviates from that found in halo stars of the same metallicity. Two notable differences are the absence of carbon-enhanced metal-poor bulge stars, and the alpha-element abundances exhibit a large intrinsic scatter and include stars which are underabundant in these typically enhanced elements.
\end{abstract}


\section{Introduction}
Studies of the most metal-poor stars have for many years provided insights into the early Universe and the formation of the Galaxy.  These stars allow us to place constraints on our understanding of the first supernovae, the early initial mass function, and the evolution of the Milky Way.  Individual metal-poor stars have led to theories on the formation of the first stars (e.g., \citealt{2012MNRAS.421.3217K}, \citealt{2012MNRAS.423L..60S}, \citealt{2014ApJ...792L..32I}) and ideas about the chemical enrichment of the galaxy thereafter (e.g., \citealt{2007ApJ...670..774N}, \citealt{2013RvMP...85..809K}, \citealt{2015ARA&A..53..631F}).

The first stars in the Universe (referred to as Population III stars) are predicted to have formed within the first few hundred million years after the Big Bang \citep[e.g.][and references therein]{2013RPPh...76k2901B}, corresponding to redshifts of $z>10$.  Until recently it was thought that these stars were all massive, and therefore short-lived \citep{2001ApJ...548...19N,2004ARA&A..42...79B}.  A lack of metals in the giant gas clouds would make the cooling needed for fragmentation into smaller clouds difficult, possibly preventing the formation of low mass stars.  However with the introduction of higher-resolution numerical simulations, it appears that accretion disc fragmentation may allow stars of around a solar mass to emerge. With that, the possibility of a Population III star surviving to the present day becomes plausible (e.g., \citealt{2011Sci...331.1040C}, \citealt{2012MNRAS.424..399G}, \citealt{2013RPPh...76k2901B}).

There have been numerous extensive searches for metal-poor stars in the Milky Way.  Surveys like the HK survey \citep{1985AJ.....90.2089B}, Hamburg-ESO survey \citep{2003RvMA...16..191C}, and SDSS/SEGUE \citep{2011A&A...534A...4C} have extended our knowledge of this area immeasurably, producing a significant number of stars with [Fe/H]\footnote{Using the standard notation, where [A/B] $\equiv log_{10}$(N$_{A}$/N$_{B}$)$_{*} - log_{10}$(N$_{A}$/N$_{B}$)$_{\sun}$, and $log_{10}\epsilon$(B) $=$ A(B) $\equiv log_{10}$(N$_{B}$/N$_{H}$) $+12.00$, for elements A and B.}$<$$-3.5$ \citep{2013ApJ...762...25N} and a few stars with [Fe/H]$<$$-5.0$  (e.g., \citealt{2002Natur.419..904C,Frebel:2005cc,2008ApJ...684..588F,2011Natur.477...67C,2014Natur.506..463K}).
In the future, large-scale spectroscopic surveys like LAMOST and 4MOST will increase the number of metal-poor stars known tenfold \citep{2015ApJ...798..110L, 2012SPIE.8446E..0TD}. These surveys have primarily targeted the Galactic halo, which is known to be on average more metal-poor than any other Galactic component. More recently this has been extended to dwarf satellite galaxies of the Milky Way (\citealt{2010A&A...524A..58T}; \citealt{2010Natur.464...72F}; \citealt{2013A&A...549A..88S}), finding stars down to [Fe/H]$=-4.0$.

According to theoretical modelling within the cold dark matter framework, however, the Milky Way halo is not the optimal place to look for the most metal-poor and oldest stars. \citet{White:2000jt} first predicted that the oldest stars in the Milky Way should mostly be in the bulge or inner halo, a conclusion which was reinforced by e.g. \citet{2007ApJ...661...10B}. \citet{2010MNRAS.401L...5S} suggested that any stars found with [Fe/H]$<$$-1$ within the inner few kpc of the Galaxy would have formed at $z>$$10$. \citet{2010ApJ...708.1398T} combined models of $\Lambda$CDM halo formation with baryonic gas budgets and star formation histories, to mimic the formation of the Milky Way.  He showed that, of all the stars with [Fe/H]$<$$-3.0$, those found in the central regions of the Galaxy were more likely to have formed before $z=15$ than in any other location.  He writes that the oldest and most metal-poor stars, which formed as early as $z\simeq20$ according to his model, ``are in the bulge, but not of the bulge". 

Despite these simulations suggesting the bulge to be the best location to find the oldest stars today, very few attempts have been made to search the bulge for metal-poor stars.  It is much easier to search the Galactic halo, where the majority of stars are of a low metallicity (the peak of the metallicity distribution function (MDF) of the halo is around [Fe/H]$=$$-1.6$; \citealt{1988AJ.....96.1908L,1991AJ....101.1865R,2009A&A...507..817S}). The bulge, on the other hand, is the most metal-rich component of the Milky Way, containing some of the most metal-rich stars known, and with an MDF ranging from [Fe/H]$\simeq$$-1.5$ to [Fe/H]$=$$+0.5$ (\citealt{2008A&A...486..177Z}; \citealt{2013MNRAS.430..836N}; \citealt{2013A&A...552A.110G}).  Indeed, the metallicity unbiased ARGOS survey \citep{2013MNRAS.428.3660F} showed that of 14,150 stars identified as lying in the bulge, only 16 had [Fe/H]$\le$$-2.0$ \citep{2013MNRAS.430..836N}. Furthermore, the bulge is also a heavily crowded region and with high extinction due to dust, making it practically very difficult to find metal-poor bulge stars. The large distance to the bulge (around 8.5\,kpc) means that only red giant stars can be targeted without amplification from microlensing \citep{2013A&A...549A.147B}. 

Very recently, the first very metal-poor stars in the bulge have been discovered. 
\citet{2013ApJ...767L...9G} found five new metal-poor stars  with [Fe/H] ranging from $-1.6$ to $-2.1$ based
on infrared spectroscopy of $\sim2,400$ bulge stars. 
Recent works combining data from near-infrared surveys with optical photometry have also started to find metal-poor bulge stars, with three discovered having $-3.0<$[Fe/H]$<-2.7$ \citep{2014ApJ...797...13S}.

Studies into the detailed chemical abundances of the bulge have in general found that the population is similar to that of the thick disc; at metallicities below about [Fe/H]$=$$-0.4$, the alpha-elements are enhanced \citep{2010A&A...513A..35A,2013A&A...549A.147B}. This enhancement implies a fast enrichment history of the bulge; before low-mass stars could enrich the environment, the bulge had already reached a metallicity of [Fe/H]$=$$-0.4$. With the discovery of lower metallicity stars, the $\alpha$ enhancement in the bulge has been probed at [Fe/H]$\approx$$-2$ -- \citet{2013ApJ...767L...9G} found that O and Mg confirmed the high-$\alpha$ trend, but that Si appeared to be lower than expected -- and [Fe/H]$\approx$$-2.5$ -- \citet{2014MNRAS.445.4241H} found some $\alpha$ enhancement, but an unusually large scatter between the stars, including sub-solar Mg values. Unfortunately, with only five and four stars respectively, neither study was able to confirm an unexpected trend.

This paper, the third in a series of papers concerning the EMBLA\footnote{In Nordic mythology, Embla was the first woman, born in the middle of the world from the remains of giants.} spectroscopic survey (Extremely Metal-poor BuLge stars with AAOmega), describes the findings of our initial observations, following the results published in \citet{2014MNRAS.445.4241H} as part of the Gaia-ESO Survey, and those published in \citet{2015Natur.527..484H} based on observations taken in 2014. These stars will crucially provide greater numbers at [Fe/H]$<$$-2$, allowing us to draw the first conclusions about the nature of the chemistry of the bulge at low metallicities. 

\section{Observations}

\subsection{Photometry from SkyMapper}

\begin{figure}
  \centering
  \includegraphics[width=0.99\columnwidth]{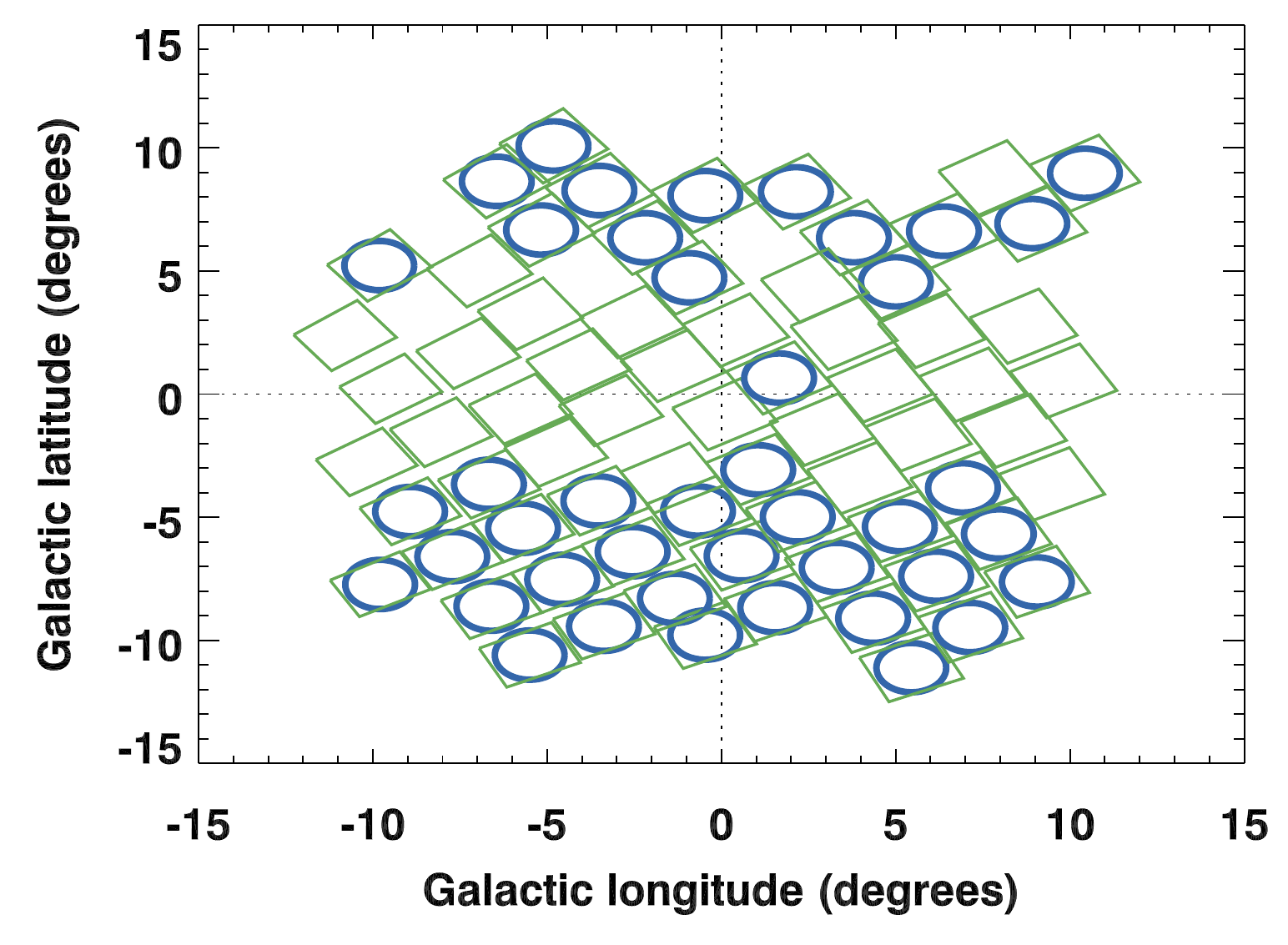}
  \caption{Positions of all the fields observed by SkyMapper in the bulge, shown as green rectangles.  The blue circles show the fields that have been followed up with spectroscopy from AAOmega.}
  \label{fig:pos}
\end{figure}

The SkyMapper telescope is a 1.3\,m telescope capable of imaging in six bandpasses with a 5.7-square-degree field of view \citep{2007PASA...24....1K}.  It has primarily been designed to perform the Southern Sky Survey, a multi-epoch photometric survey of the whole of the southern sky, which commenced regular science operations in 2014.  We have acquired complete coverage of the bulge with SkyMapper, taken during telescope commissioning during 2012-2014.  The distribution of these fields is shown in Figure \ref{fig:pos}.  Each bulge field contains on the order of $10^{6}$ stars, ranging roughly from 12th to 19th magnitude in the Str\"omgren $v$ band.

\begin{figure}
  \centering
  \includegraphics[width=0.99\columnwidth]{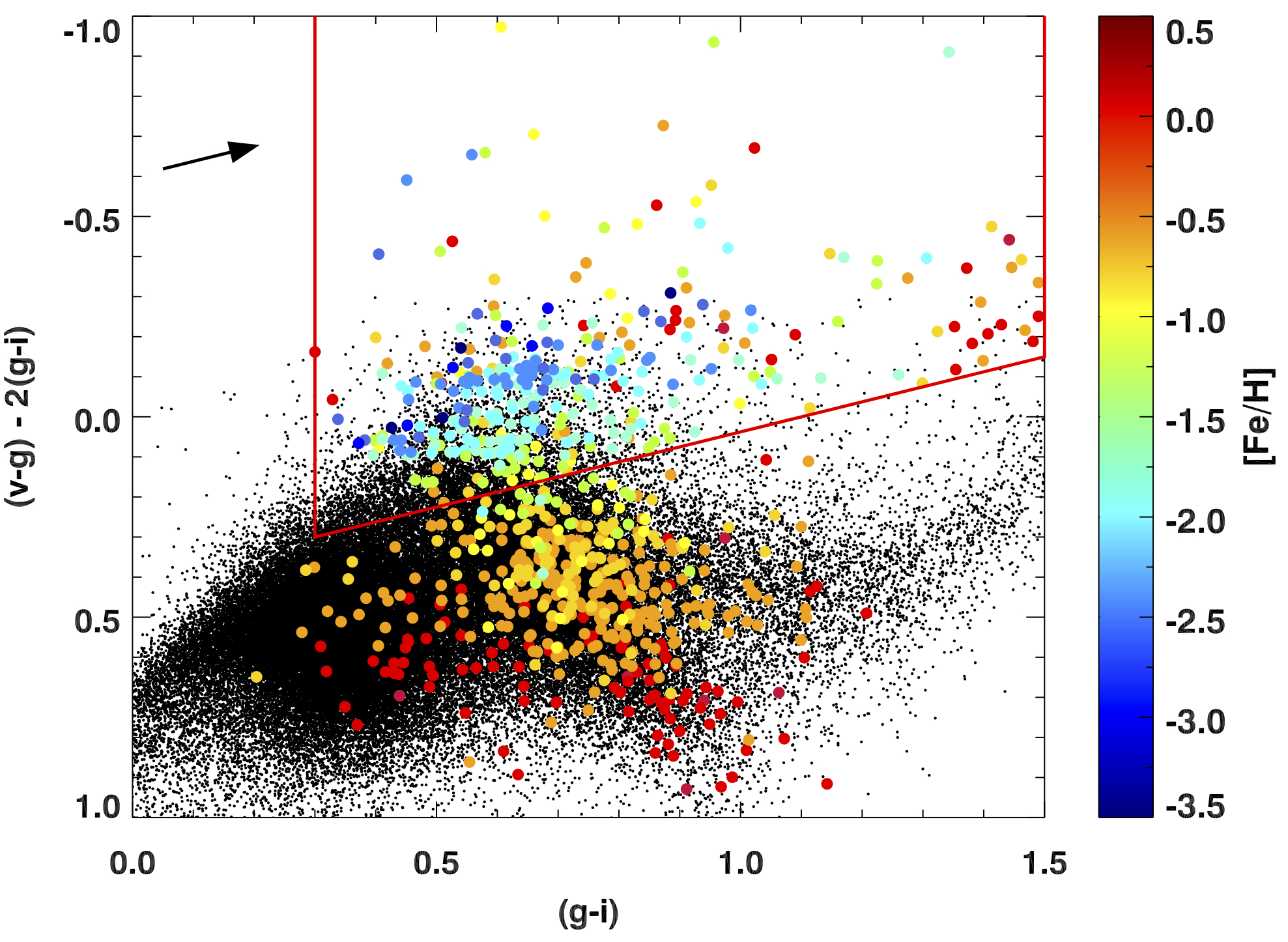}
  \caption{Two-colour plot using the $g$, $v$, and $i$ bands of SkyMapper to demonstrate the metallicity dependence on the $(v-g)-2(g-i)$ colour.  The coloured circles are data taken from both EMBLA and the ARGOS survey \citep{2013MNRAS.430..836N}, with [Fe/H] determined spectroscopically.  The red trapezium shows our selection region for metal-poor candidates.  The arrow represents the mean reddening vector in this field, $E(B-V)$=0.17 \citep{2011ApJ...737..103S}.}
  \label{fig:photometry}
\end{figure}

The filters in SkyMapper have been chosen specifically to enhance important spectral features in both stellar and extra-galactic research \citep{2011PASP..123..789B}; in our case in particular the $v$ band (when combined with the $g$ and $i$ bands) provides an important stellar metallicity indicator. We first select stars on the giant branch of the bulge from a $g-i$, $g$ colour-magnitude diagram.  This is necessary to limit our selection to stars that are in the bulge, and not foreground dwarfs.  Then we move on to the metallicity selection, plotting $(v-g) - 2(g-i)$ against $g-i$ to create a selection box. In Figure \ref{fig:photometry} we show an example two-colour diagram, revealing its powerful ability to identify low metallicity stars. We have overlaid the SkyMapper photometry with spectroscopic [Fe/H] data taken from both a metallicity unbiased ARGOS field, and an EMBLA field, centred at ($l, b$)$=$($0, -10$). We chose the selection regions in each field (shown as the red box in Figure \ref{fig:photometry}) by selecting the 700 stars with the lowest $(v-g) -2(g-i)$ values, adjusting this index in each field to account for reddening. From the selection region, a box of the first 200 stars from the top downwards were identified as the highest priority candidates, followed by a box containing 500 lower priority candidates from which $\sim150$ are chosen at random at the fibre configuration stage of the spectroscopic follow-up. This second selection is designed to provide us with a random sample of the metal-poor stars in the field, from which we aim to recreate the tail end of the metallicity distribution function (MDF) of the bulge to be presented in a later paper.

\subsection{Medium resolution spectroscopy with AAOmega on the AAT}
\begin{figure}
  \centering
  \includegraphics[width=0.99\columnwidth]{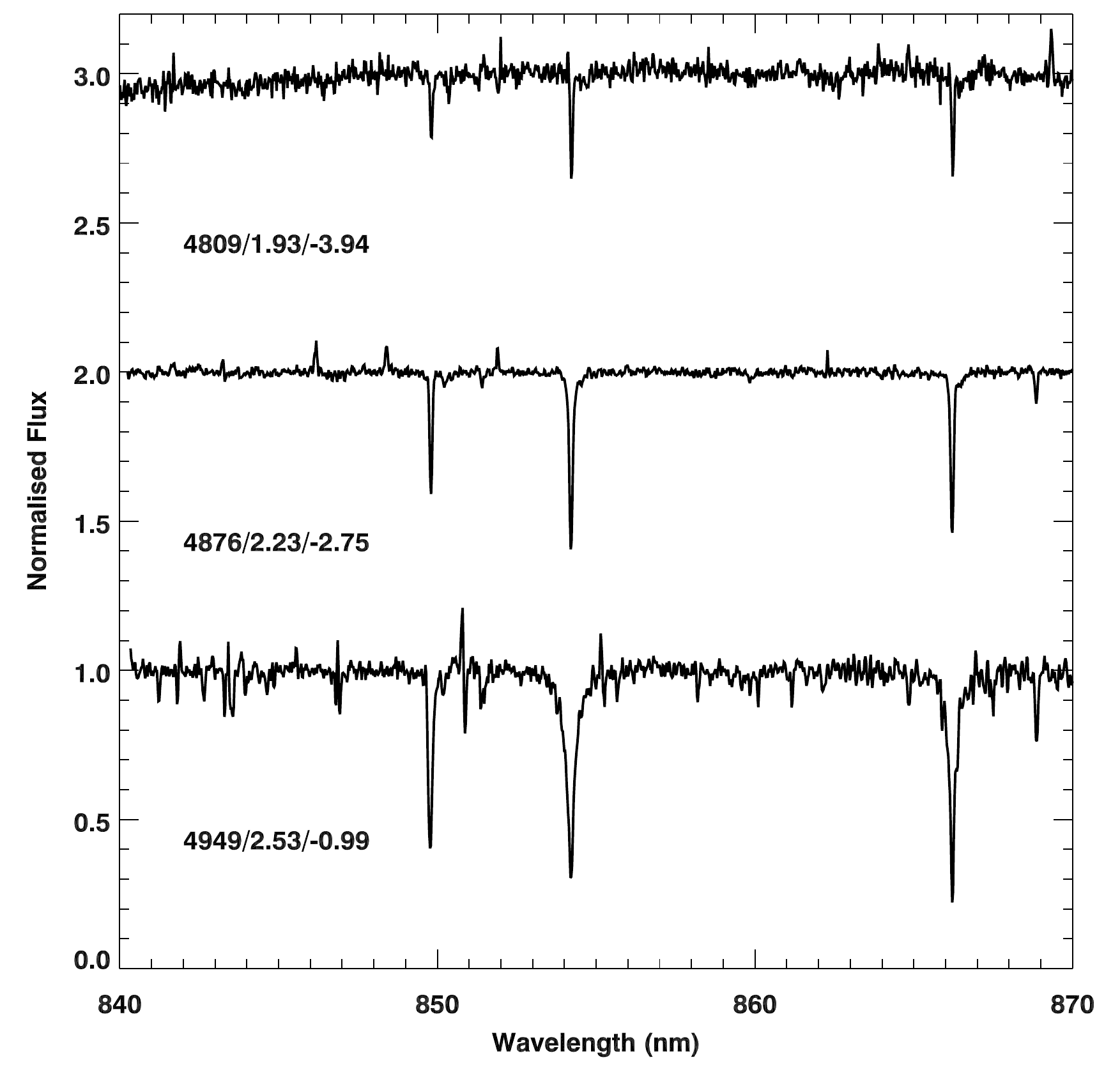}
  \caption{The red-arm spectra showing the Ca\,\textsc{ii} triplet (R=10,000) of three of the stars observed with AAOmega.  The three stars have been chosen from the same field ($l, b$)$=$($-1.5$, $-8.8$) and with a range of metallicities to demonstrate the notable difference in the spectra with varying metallicity.  The stellar parameters ($T_{\rm{eff}}$,  $\log{g}$, [Fe/H]) are labelled underneath each spectrum.}
  \label{fig:aatspec}
\end{figure}

With the capability of selecting so many candidate metal-poor stars, an efficient means to spectroscopically confirm their metal-poor nature is necessary. The AAOmega spectrograph combined with the 2dF fibre positioner on the Anglo-Australian Telescope (AAT) \citep{2006SPIE.6269E..14S} provides spectra of up to 392 stars at once with a circular field-of-view of $2\degr$ diameter.  Over 24.5 nights on the AAT between 2012 and 2014 we have observed more than 14,000 stars in the fields shown in Figure \ref{fig:pos}.  All observations were taken using the 1700D grating for the red arm, and the 580V grating for the blue arm, which provides a spectral resolving power of 10,000 in the Ca II triplet region and of 1,300 over 370 - 580\,nm.  The data were reduced using the 2dfdr pipeline\footnote{http://www.aao.gov.au/science/software/2dfdr} (version 5.39), and examples of reduced spectra from field 2156 (($l, b$)$=$($-1.5, -8.3$)) can be seen in Figure \ref{fig:aatspec}, showing a range of metallicities.  

\begin{figure}
  \centering
  \includegraphics[width=0.99\columnwidth]{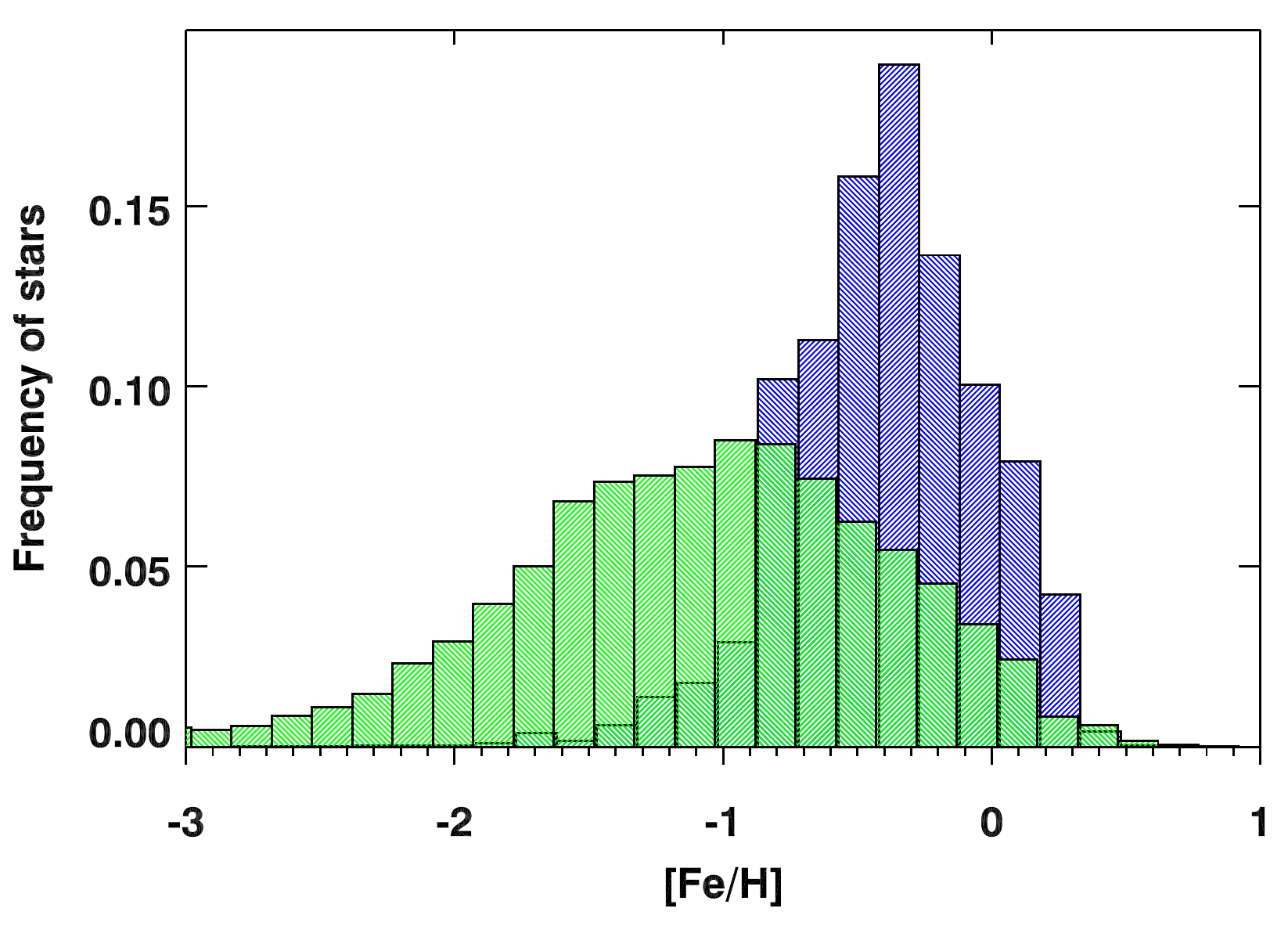}
  \caption{Raw MDF of the first 9,000 spectra from the EMBLA Survey (green) observed in 2012 and 2013, compared to the MDF of the ARGOS bulge survey (blue).  Both are normalised to have the same area, and the EMBLA histogram has been truncated at [Fe/H]$=$$-3.0$  due to a combination of reduction and analysis issues producing some false positives at low metallicity.}
  \label{fig:histogram}
\end{figure}

The AAOmega spectra have been analysed using \texttt{sick} \citep{2016ApJS..223....8C}, a Python code that forward models spectroscopic data from which we can ascertain the standard astrophysical parameters of the stars: $v_{\rm{rad}}$, $T_{\rm{eff}}$,  $\log{g}$, [Fe/H], and [$\alpha$/Fe]. We interpolated synthetic spectra from the AMBRE grid \citep{2012A&A...544A.126D}.  From these results, we have calculated the raw MDF of the data (Fig. \ref{fig:histogram}), which demonstrates the overall success of the photometric selection process.  When compared to all bulge stars from the ARGOS survey (shown in blue), the peak of our MDF is $\sim0.6$\,dex lower, and we have a significant number of stars reaching down to the lowest metallicities. The fields observed span a range of locations in the bulge (as seen in Fig. \ref{fig:pos}), which will allow us to complete a detailed breakdown of the metal-poor stellar population across the bulge.  We note that as expected, the SkyMapper photometric selection of metal-poor stars is less successful in the heavily and differentially reddened bulge region than for the halo \citep{2015ApJ...807..171J}. On the other hand, however, the spectroscopic confirmation stage is far more efficiently carried out using the high multiplexing of AAOmega.

\subsection{High resolution spectroscopy with MIKE on Magellan}
In order to discover the detailed chemical composition of some of the most metal-poor stars found in the survey, we have observed at both Magellan \citep{2015Natur.527..484H} and the VLT \citep{2014MNRAS.445.4241H} over the course of the three years of the survey.  This paper focuses on high-resolution data taken at Magellan in 2012, immediately after the first fields were observed on the AAT.  As they were observed early in the course of the survey, these are not the most metal-poor stars we have discovered, rather a range of stars with metallicities originally estimated as [Fe/H]$<$$-2$.

\begin{table*}
\begin{minipage}{180mm}
\centering
\caption{Details of the observations of the 10 stars. The SkyMapper naming convention is SMSS J(RA2000)+(Dec2000).}
\label{table:obs}
\begin{tabular}{lrrrrrrr}
\hline
Star (SMSS) & $l$ & $b$ & Date Observed & Exposure & Wavelength & Slit width & S/N per pixel \\
 & ($\degr$) & ($\degr$) & (2012) & time (s) & range (nm) & ('') & at 450nm \\
\hline
 J182637.10-342924.2 & -0.68 & -10.31 & April 28 & 14,400 & 340-890 & 0.35 & 35 \\
 J182600.09-332531.0 & 0.24 & -9.72 & May 15 & 1,200 & 333-941 & 1.0 & 20 \\
 J182601.24-332358.3 & 0.26 & -9.72 & May 15 & 2,700 & 333-941 & 1.0 & 20 \\
 J182753.81-334607.7 & 0.10 & -10.23 & May 15 & 1,200 & 333-941 & 1.0 & 13 \\
 J183000.36-333919.3 & 0.40 & -10.58 & May 15 & 1,350 & 333-941 & 1.0 & 24 \\
 J182922.48-335559.4 & 0.09 & -10.58 & May 15 & 3,600 & 333-941 & 1.0 & 28 \\
 J182930.47-335958.3 & 0.04 & -10.63 & May 15 & 1,850 & 333-941 & 1.0 & 17 \\
 J183225.29-334938.4 & 0.46 & -11.10 & May 15 & 1,350 & 333-941 & 1.0 & 16 \\
 J183128.71-341018.4 & 0.06 & -11.07 & May 15 & 900 & 333-941 & 1.0 & 14 \\
 J182948.48-341053.9 & -0.10 & -10.77 & June 25 & 3,600 & 370-890 & 1.0 & 26 \\
\hline
\end{tabular}
\end{minipage}
\end{table*}

Ten stars were observed using the MIKE high-resolution spectrograph \citep{2003SPIE.4841.1694B} on Magellan's 6.5m Clay telescope.  The observations took place between April and June of 2012, and all make use of the full wavelength coverage offered by MIKE, with the spectra covering (as a minimum) 370-890 nm. All except one star (SMSS J182637.10-342924.2) were configured with a 1.0" slit, resulting in resolving powers of 22,000 in the blue and 28,000 in the red, and were binned by two in both the spatial and spectral directions. SMSS J182637.10-342924.2 was instead observed in April as part of a different set of observations, where a slit with width of 0.35" and no spatial or spectral binning, was used.  This provided resolving powers of 83,000 and 65,000, in the blue and red respectively.  We note that although the higher resolving power provides extra detail in our spectra, the lower resolving powers are sufficient to be able to measure the elements of relevance at these modest signal-to-noise values of metal-poor stars. Details of the exact S/N acquired (after binning), along with exposure times and wavelength range can be found in Table \ref{table:obs}.

\section{Analysis}

\subsection{Data Reduction and Radial Velocities}
The high-resolution spectra were all reduced with the CarPy data reduction pipeline\footnote{http://code.obs.carnegiescience.edu/mike}, version 2014-04-24 \citep{2003PASP..115..688K}. The spectra were normalised using \textsc{smh} \citep{2014arXiv1405.5968C}, a user interface that combines the normalisation, doppler correction, equivalent width measurement, stellar parameter determination, and chemical abundance calculation of a high-resolution spectrum. The orders were fit with a cubic spline, with prominent lines and band heads masked out. Radial velocities for the stars were determined using both \textsc{iraf} and \textsc{smh}.  Both methods make use of cross-correlation with the spectrum of a known metal-poor subgiant (HD140283), and their respective velocities were found to have on average less than $1.0$\,km\,s$^{-1}$ difference. The \textsc{smh} values are used throughout, and converted to heliocentric radial velocities using \textsc{iraf}. The radial velocities are all single epoch measurements, leaving open the possibility that some fraction of the stars are in binary systems.

\subsection{Atmospheric Parameters}

\subsubsection{Equivalent Widths}
The strengths of atomic absorption lines were measured in all 10 stars using a line list compiled for the EMBLA Survey, listed in Table \ref{table:atom} along with the adopted atomic data.  This line list was extracted from the Gaia-ESO line list (\citealt{2015PhyS...90e4010H}; Heiter et al. in prep.), specifically utilising the best lines available in metal-poor stars, and supplemented with lines outside the wavelength regions of Gaia-ESO with lines primarily from \citet{2013ApJ...762...25N}.  In particular, $66$ Fe\,\textsc{i} lines and $24$ Fe\,\textsc{ii} lines have been included, in order to robustly measure [Fe/H] and other atmospheric parameters.  The lines were measured automatically using \textsc{smh}, which determines the local continuum and then iteratively fits a Gaussian profile to the line. Line broadening parameters were taken from \citet{2000A&AS..142..467B} and \citet{2005A&A...435..373B} where possible, otherwise treated using the Uns\"old approximation \citep{1955psmb.book.....U}. All lines were then checked by eye, removing any spurious results, and rejecting any line with an equivalent width greater than $120$ m\AA\ in order to restrict our equivalent width measurements to the linear part of the curve of growth. For the most metal-poor star in the sample (SMSS J182601.24-332358.3), $18$ Fe\,\textsc{i} lines and $8$ Fe\,\textsc{ii} lines were measurable. In some cases, where there are no other lines available for a particular element, lines stronger than $120$ m\AA\  have been used. The equivalent widths measured for all ten stars are given in Table \ref{table:eqwidths}.

\begin{table}
\centering
\caption{Atomic data. The full table is published online, part of the table is shown here to demonstrate the content.}
\label{table:atom}
\begin{tabular}{lrrrr}
\hline
Ion & $\lambda$ (nm) & $\chi$ (eV) & $\log{gf}$ & Reference\\
\hline
Li\,\textsc{i} & 670.776 & 0.000 & 0.174 & 1. \\
O\,\textsc{i} & 630.030 & 0.000 & -9.715 & 2. \\
O\,\textsc{i} & 636.378 & 0.000 & -10.190 & 2. \\
Na\,\textsc{i} & 588.995 & 0.000 & 0.108 & 3. \\
Na\,\textsc{i} & 589.592 & 0.000 & -0.194 & 3. \\
Na\,\textsc{i} & 818.326 & 2.102 & 0.237 & 4. \\
Na\,\textsc{i} & 819.482 & 2.104 & 0.492 & 4. \\
... & ... & ... & ... \\
\hline
\end{tabular} \\
1. \citet{Lithium}, 2. \citet{Oxygen},
3. \citet{Sodium}, 4. \citet{NIST10}, 5. \citet{GARZ},
6. \citet{K07}, 7. \citet{BL}, 8. \citet{Potassium},
9. \citet{SG}, 10. \citet{Calcium},
11. \citet{SR}, 12. \citet{S},
13. \citet{Calcium2},
14. \citet{LD}, 15. \citet{K09}, 16. \citet{LGWSC},
17. \citet{BHN},
18. \citet{K10},
19. \citet{WLSC},
20. \citet{SLS},
21. \citet{DLSSC},
22. \citet{BWL},
23. \citet{BK},
24. \citet{BKK},
25. \citet{GESB79b},
26. \citet{Iron1},
27. \citet{Iron2},
28. \citet{FMW},
29. \citet{GESB82c},
30. \citet{GESHRL14},
31. \citet{BIPS},
32. \citet{MRW},
33. \citet{GESB82d},
34. \citet{Iron3},
35. \citet{RU},
36. \citet{WLSCow},
37. \citet{K08},
38. \citet{KR},
39. \citet{Copper},
40. \citet{Zinc1},
41. \citet{Zinc2},
42. \citet{HLGBW},
43. \citet{LNAJ},
44. \citet{LBS},
45. \citet{LWHS}
\end{table}

\begin{table*}
\begin{minipage}{180mm}
\centering
\caption{The measured equivalent widths of the 10 stars for each spectral line. Lines from Table \ref{table:atom} that are missing here were not measured in any of these stars. The full table is published online, part of the table is shown here to demonstrate the content.}
\label{table:eqwidths}
\begin{tabular}{lrrrrrrrrr}
\hline
Ion & $\lambda$ (nm) & J182637.10 & J182600.09 & J182601.24 & J182753.81 & J183000.36 & J182922.48 & J182930.47 &...\\
 & & -342924.2 & -332531.0 & -332358.3 & -334607.7 & -333919.3 & -335559.4 & -335958.3 & \\
\hline
O\,\textsc{i} & 630.030 & - & - & - & - & - & - & 19.4 & ... \\
O\,\textsc{i} & 636.378 & - & - & - & - & - & - & 10.7 & ... \\
Na\,\textsc{i} & 588.995 & 193.3 & 213.7 & 199.0 & 199.9 & 176.7 & 179.4 & 248.0 & ... \\
Na\,\textsc{i} & 589.592 & 169.7 & 203.0 & - & - & 151.5 & 128.4 & 201.4 & ... \\
Na\,\textsc{i} & 818.326 & - & - & - & - & - & - & - & ... \\
Na\,\textsc{i} & 819.482 & 62.1 & - & - & 50.9 & - & - & - & ... \\
... & ... & ... & ... & ... & ... & ... & ... & ... & ... \\
\hline
\end{tabular}
\end{minipage}
\end{table*}

\subsubsection{Effective Temperature}
\begin{figure*}
\begin{minipage}{180mm}
  \centering
  \includegraphics[width=0.85\columnwidth]{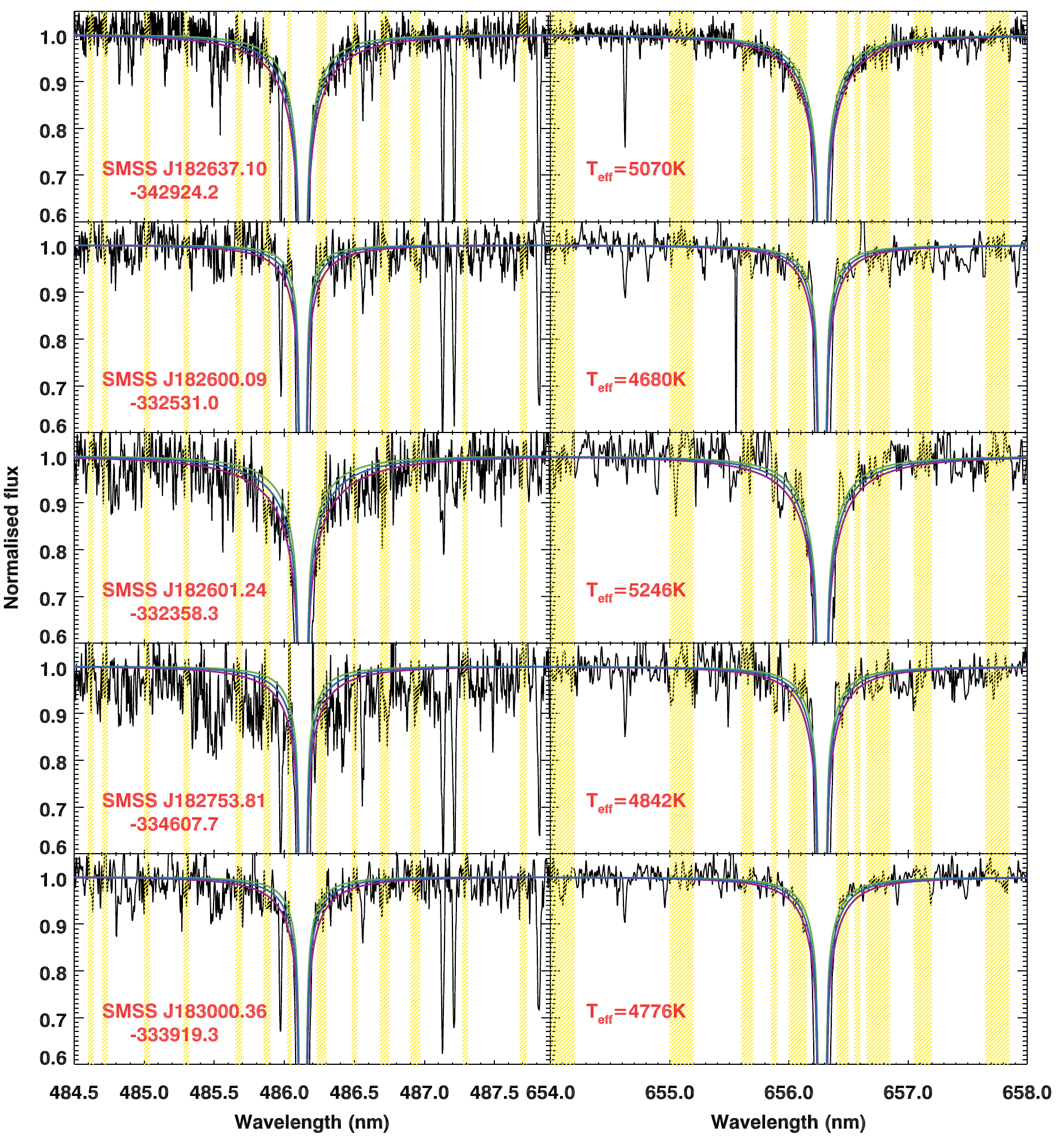}
  \caption{The spectra of the first five stars at both the H$\beta$ and H$\alpha$ lines, from which a synthetic spectrum was fitted to derive the effective temperatures. The yellow shaded regions show the spectral windows used for the fit. Three synthetic spectra have been overplotted; one with the fitted temperature (blue), one with $T_{\rm{eff}}$ $+160$\,K (purple), and one with  $T_{\rm{eff}}$ $-160$\,K (green). The fits for SMSS J182600.09-332531.0 were not used to derive the $T_{\rm{eff}}$ of this star, see text for details.}
  \label{fig:hlines1}
\end{minipage}
\end{figure*}

\begin{figure*}
\begin{minipage}{180mm}
  \centering
  \includegraphics[width=0.85\columnwidth]{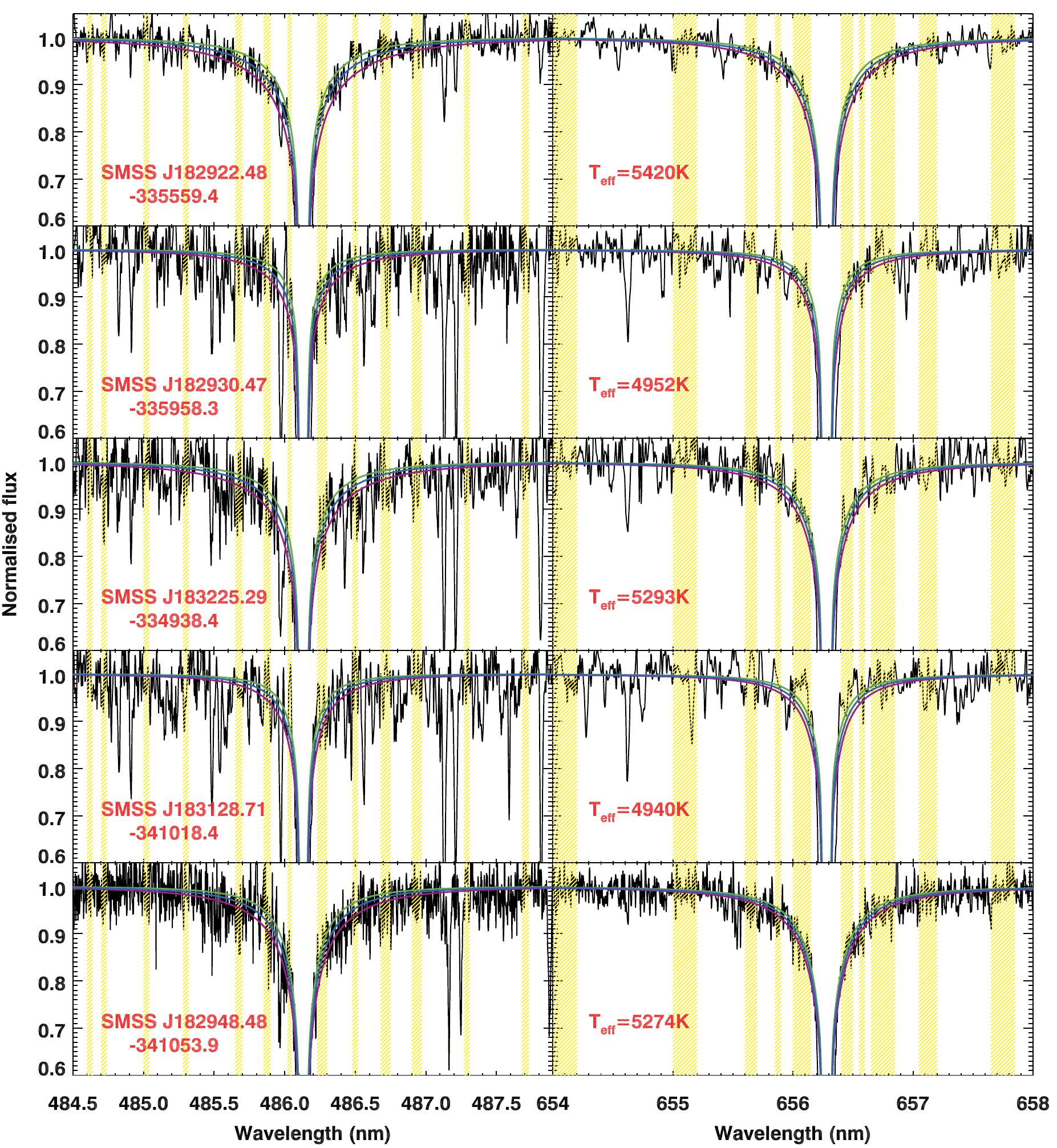}
  \caption{Same as Fig. \ref{fig:hlines1} for the remaining five stars. The fits for SMSS J183128.71-341018.4 were not used to derive the $T_{\rm{eff}}$ of this star, see text for details.}
  \label{fig:hlines2}
\end{minipage}
\end{figure*}

The atmospheric parameters for the stars were calculated using an iterative process, where the initial parameters were taken from those derived from the low-resolution spectra. $T_{\rm{eff}}$ values were found for each star by interpolating between a grid of precomputed synthetic H$\alpha$ and H$\beta$ lines (\citealt{2002A&A...385..951B}\footnote{http://www.astro.uu.se/~barklem/data.html}, using both the G and K giant grids with [$\alpha$/Fe]$=$$0$, and the metal-poor giant grids with [$\alpha$/Fe]$=$$0.4$), and matching to suitable wavelength regions of the observed spectra using a $\chi^{2}$ minimisation (Figures \ref{fig:hlines1} and \ref{fig:hlines2}). This method has been preferred over the excitation potential balance method, due to the latter producing significantly lower temperatures in metal-poor stars than any other method \citep{2004A&A...416.1117C,2008ApJ...681.1524L, 2013ApJ...769...57F}.  Alternate temperatures were also derived using excitation balance, followed by an empirical correction of +325\,K, decided upon after analysing both the metal-poor Gaia benchmark stars \citep{2015A&A...582A..49H} and the larger sample of stars in \citet{2015Natur.527..484H} (25 stars in total). In general, these alternate temperatures were used solely to confirm that our temperatures from the Balmer lines were accurate, however due to the uncertain normalisation of the hydrogen wings in two of the stars (SMSS J182600.09-332531.0 and SMSS J183128.71-341018.4, both of which have quite low temperatures, which results in smaller hydrogen lines), the corrected excitation balance $T_{\rm{eff}}$ values were used instead for those two stars.

\subsubsection{Surface Gravity, Metallicity, and Microturbulence}
After the effective temperatures were determined, these values were entered into \textsc{smh} (which provides a user interface for the 1D LTE stellar synthesis software \textsc{moog} \citep{2012ascl.soft02009S}), and the remaining atmospheric parameters were measured, using the 1D MARCS model atmospheres \citep{2008A&A...486..951G}, that have an $\alpha$-enhancement of [$\alpha$/Fe]$=$$0.4$. The $\log{g}$ values were determined first by forcing the mean Fe\,\textsc{i} and Fe\,\textsc{ii} abundances to be equal (within $0.01$\,dex).  Similarly, the microturbulence was calculated by forcing a zero gradient between the Fe\,\textsc{i} abundances and the reduced equivalent widths, with a maximum value set at $\xi_{t}=2.5$ km\,s$^{-1}$. The adopted [Fe/H] values are the mean Fe\,\textsc{ii} abundances from the line measurements. Non-LTE effects in the measurements of the Fe\,\textsc{i} lines were considered using the calculations of \citet{2012MNRAS.427...50L}. As our stars are both metal-poor and giants, the non-LTE corrections were typically quite large: the average correction applied to the sample was 0.14\,dex. These new parameters were then used to re-calculate the effective temperatures, and the process repeated until the process had converged. The final parameters are listed in Table \ref{table:params}.

\begin{table*}
\begin{minipage}{180mm}
\centering
\caption{Stellar parameters for the 10 bulge stars.}
\label{table:params}
\begin{tabular}{lrrrrrr}
\hline
Star (SMSS) & $V_{GC}$ (km\,s$^{-1}$)$^\ast$ & d$_{\odot}$ (kpc) & $T_{\rm{eff}}$ (K) & $\log{g}$ (cgs) & [Fe/H]  (dex) & $\xi_{t}$ (km\,s$^{-1}$) \\
\hline
 J182637.10-342924.2 & -44.7 & 8.1 $\pm2.3$ & 5070 $\pm160$ & 2.50 $\pm0.13$ & -1.97 $\pm0.08$ & 1.3 $\pm0.2$\\
 J182600.09-332531.0 & -213.2 & 10.6 $\pm3.1$ & 4680 $\pm160$ & 1.36 $\pm0.13$ & -2.53 $\pm0.08$ & 2.4 $\pm0.2$\\
 J182601.24-332358.3 & -323.3 & 10.5 $\pm2.9$ & 5246 $\pm160$ & 1.65 $\pm0.12$ & -2.83 $\pm0.11$ & 2.5 $\pm0.2$\\
 J182753.81-334607.7 & -270.3 & 12.3 $\pm3.4$ & 4842 $\pm160$ & 1.53 $\pm0.12$ & -2.31 $\pm0.06$ & 2.5 $\pm0.2$\\
 J183000.36-333919.3 & 124.2 & 10.4 $\pm2.9$ & 4776 $\pm160$ & 1.55 $\pm0.12$ & -2.63 $\pm0.07$ & 2.5 $\pm0.2$\\
 J182922.48-335559.4 & -54.0 & 9.9 $\pm2.7$ & 5420 $\pm160$ & 1.94 $\pm0.12$ & -2.77 $\pm0.08$ & 2.5 $\pm0.2$\\
 J182930.47-335958.3 & -218.8 & 16.9 $\pm4.9$ & 4952 $\pm160$ & 1.25 $\pm0.13$ & -1.97 $\pm0.12$ & 2.3 $\pm0.2$\\
 J183225.29-334938.4 & -147.4 & 9.2 $\pm2.6$ & 5293 $\pm160$ & 2.35 $\pm0.12$ & -1.74 $\pm0.09$ & 1.8 $\pm0.2$\\
 J183128.71-341018.4& -35.0 & 3.3 $\pm1.0$ & 4940 $\pm160$ & 2.15 $\pm0.13$ & -1.83 $\pm0.10$ & 2.0 $\pm0.2$\\
 J182948.48-341053.9 & -71.7 & 6.0 $\pm2.3$ & 5274 $\pm160$ & 2.82 $\pm0.13$ & -2.47 $\pm0.11$ & 2.2 $\pm0.2$\\
\hline
\end{tabular}
\end{minipage}
$^\ast$ Galactocentric velocity, calculated using equation \ref{eq:gal}.
\end{table*}

\subsubsection{Uncertainties in the Parameters}
The statistical uncertainties of the $\chi^{2}$ minimisation employed in the determination of the effective temperatures were typically 125\,K for the stars. The uncertainties are also dominated by systematic errors, which have been estimated to be $\sim100$\,K \citep{2002A&A...385..951B}, when combined in quadrature, this gives a total temperature uncertainty of 160\,K. The $\log{g}$ uncertainties are assumed to be composed of the difference in $\log{g}$ when the Fe\,\textsc{ii} line abundances are altered by their standard error, and an added term of 0.12\,dex, half the size of the average $\log{g}$ correction from non-LTE effects, to encompass the uncertainty in that correction. For microturbulence, an uncertainty of 0.2\,km\,s$^{-1}$ is assumed throughout.  The metallicity uncertainties were calculated by summing in quadrature the effects of the uncertainties from $T_{\rm{eff}}$ (average of 0.02), $\log{g}$ (average of 0.05), and $\xi_{t}$ (average of 0.04), as well as the standard error of the Fe\,\textsc{ii} abundance (average of 0.06).

\subsection{Chemical Abundances}
It was possible to measure the chemical abundances of up to 21 elements additional to Fe from the spectra. Equivalent widths were measured for 19 elements, using the lines listed in Table \ref{table:atom}. These elements covered include light (O, Na, Al, K), alpha (Mg, Si, Ca, Ti), iron group (Sc, Mn, Cr, Co, Ni), and neutron-capture (Zn, Sr, Y, Zr, La, Eu). C abundances were derived from synthesising the CH bands at 431.3\,nm (430.5 - 431.9\,nm) and 432.3\,nm (431.9 - 432.9\,nm) \citep{2014A&A...571A..47M}. The Ba lines were synthesised in order to account for isotopic and hyperfine splitting, again taking atomic data from the Gaia-ESO line list. Lines of Sc, Mn, and Co were also synthesised to account for hyperfine structure, however the lines were very small and the difference in derived abundance between synthesis and equivalent width measurement was negligible.

The inferred abundances are listed in the form of [X/Fe] in Table \ref{table:abunds}, calculated relative to the solar values given in \citet{2009ARA&A..47..481A}. The Fe abundances used in these calculations were the Fe\,\textsc{ii}-based values, apart from for the iron-peak elements Cr, Mn, Co, Ni, and Zn. These neutral species behave more similarly to Fe\,\textsc{i}, for example in terms of their non-LTE effects (e.g., \citealt{2005ARA&A..43..481A}, \citealt{2012MNRAS.427...27B}), and dependence on stellar parameters, especially $\log{g}$. Their abundances are therefore given relative to the (uncorrected for NLTE) Fe\,\textsc{i} abundances. The uncertainties given for [X/Fe] are calculated by summing in quadrature the offsets due to the uncertainties in $T_{\rm{eff}}$, $\log{g}$, $\xi_{t}$, and [Fe/H] (which were $\sim0.05$\,dex), as well as the standard error across the individual line abundances. For elements where only one line was measurable, 0.10\,dex has been used to represent this standard error. Besides Fe, we have implemented line-by-line non-LTE corrections for Na \citep{2011A&A...528A.103L}, Mg, and Ca (Lind et al. in preparation).

\subsection{Distances to the stars}
Distances to the stars have been estimated by calculating the distance modulus from the absolute and apparent magnitudes. The SkyMapper photometry that we used in the initial selection was taken early on in the telescope's commissioning period, and as such was not calibrated to other magnitude systems. Therefore we chose to use 2MASS $J$, $H$, and $K_{S}$ bands, from which we reconstructed the apparent bolometric flux of the stars in magnitudes, fitting synthetic model fluxes to the stellar parameters of the stars and reddening values taken from \citet{2011ApJ...737..103S}. As the majority of the ten stars analysed here lie more than 10$\degr$ away from the plane, they are not covered by the more recently published bulge reddening maps such as OGLE \citep{2013ApJ...769...88N} and VVV \citep{2011A&A...534A...3G} and instead the \citet{2011ApJ...737..103S} reddening map has been employed. The process of reconstructing the bolometric fluxes is documented in \citet{2006MNRAS.373...13C} and \citet{2012ApJ...761...16C}. Absolute magnitudes were calculated from the Stefan-Boltzmann relation, specifically using 
\begin{equation}
\centering
M_{bol}=M_{bol,\odot}-2.5\log{\frac{L_{*}}{L_{\odot}}},
\end{equation}
where
\begin{equation}
\centering
{L}_{*}={\frac{4 \pi \sigma {T_{\rm{eff}}}^{4}{M}_{*}G}{10^{\log{g_{*}}}}}.
\end{equation}
We have assumed a stellar mass of $M_{*}=0.8\pm0.2\,M_{\odot}$ throughout, appropriate for the very old, metal-poor stars studied here. The absolute ($M_{bol}$) and apparent ($m_{bol}$) bolometric fluxes were then used to compute the distance ($d$) using
\begin{equation}
\centering
\log_{10}{d}=1+\frac{m_{bol}-M_{bol}}{5}.
\end{equation} 

\section{Analysis of previously studied stars}
This study follows on from the chemical abundance results published in the first paper of the EMBLA Survey \citep{2014MNRAS.445.4241H}. It is important to include those stars published in \citet{2014MNRAS.445.4241H} as part of the Gaia-ESO Survey (GES, \citealt{2012Msngr.147...25G}; \citealt{2013Msngr.154...47R}) when analysing the sample described here for a fair comparison. Full details of the observations and original analysis of the GES stars can be found in \citet{2014MNRAS.445.4241H} and corresponding survey papers, but we will briefly outline it here. A further 23 stars have been analysed in \citet{2015Natur.527..484H}, and a chemical analysis of those results will be the subject of a forthcoming paper. 

\begin{table}
\centering
\caption{Details of the four stars originally published in \citet{2014MNRAS.445.4241H}.}
\label{table:params2}
\begin{tabular}{lrrrrr}
\hline
Star (SMSS) & l ($\degr$) & b ($\degr$) & $V_{GC}$  & S/N$^*$  \\
 & & &(kms$^{-1}$)&  & \\
\hline
 J182153.85-341018.8 & 359.2 & -9.3 & -237.68 &  73 \\
 J183617.33-270005.3 & 7.1 & -8.9 & -129.48 &  37 \\
 J175510.50-412812.1 & 350.2 & -8.0 & -48.28 &  14  \\
 J175652.43-413612.8 & 350.2 & -8.4 & 216.46 & 14 \\
\hline
\end{tabular}
$^\ast$ Median S/N per pixel calculated across total wavelength range.
\end{table}

The Gaia-ESO Survey is a public spectroscopic survey taking place on the VLT, aiming to observe approximately 100,000 stars in the Galaxy.  All major components of the Milky Way are being covered, including the bulge. The stars are observed using the FLAMES multi-object spectrograph  \citep{2000SPIE.4008..534D}, which combines both GIRAFFE and UVES to observe up to 138 targets at once. As part of a collaboration between the SkyMapper and GES teams, extremely metal-poor targets are observed with the UVES fibres in both halo and bulge fields. In 2012, six bulge targets were observed for the EMBLA Survey, resulting in spectra covering the region of 480-680\,nm at a resolving power of 47,000. Unfortunately, two of the spectra had an average S/N lower than 10, and were unable to be analysed, but four spectra were processed (Table \ref{table:params2}). The data were reduced and normalised along with the rest of the survey data \citep{2014A&A...565A.113S}, but the analysis was separate from the standard GES UVES analysis \citep{2014A&A...570A.122S} due to the metal-poor nature of the stars. Four of the analysis nodes provided accurate atmospheric parameters for the Gaia metal-poor benchmark stars \citep{2014A&A...564A.133J}, and so their parameters for the EMBLA stars were combined in a weighted average along with our own analysis. This analysis was the alternate method using excitation balance along with an empirically calibrated offset, as described in Section 3.2.3. The uncertainties were calculated as the standard errors between the five different analysis methods.

In order to compare the results from these stars with the stars observed with Magellan, we have also analysed them using the same methods described in Section 3. Unfortunately the S/N in two of these stars was too low to ascertain sensible $T_{\rm{eff}}$ estimates. Instead for these two stars, we again used the alternate temperatures derived from excitation-balance, offset by +325\,K in a similar determination to that of SMSS J182600.09-332531.0 and SMSS J183128.71-341018.4 using the Magellan spectra described above. The parameters from both Gaia-ESO and this paper are compared in Table \ref{table:compare}, and the new abundances calculated are listed in Table \ref{table:abunds}.

\begin{table*}
\begin{minipage}{180mm}
\centering
\caption{Comparison of the new parameters with the GES recommended parameters.}
\label{table:compare}
\begin{tabular}{llrrrrr}
\hline
Star  (SMSS) & Analysis & d$_{\odot}$ (kpc) & $T_{\rm{eff}}$ (K) & $\log{g}$ (cgs) & [Fe/H] (dex) & $\xi_{t}$ (km\,s$^{-1}$) \\
\hline
J182153.85-341018.8 & Gaia-ESO & 7.0 $\pm3.2$ & 4947 $\pm85$ & 1.41 $\pm0.49$ & -2.60 $\pm0.31$ & 2.3 $\pm0.2$ \\
 & This paper & 7.6 $\pm2.1$ & 4896 $\pm160$ & 1.33 $\pm0.12$ & -2.51 $\pm0.07$ & 1.9 $\pm0.2$ \\
J183617.33-270005.3 & Gaia-ESO & 5.3 $\pm1.9$ & 4926 $\pm137$ & 1.97 $\pm0.35$ & -2.72 $\pm0.28$ & 2.4 $\pm0.2$ \\
 & This paper & 5.5 $\pm1.5$ & 4842 $\pm160$ & 1.93 $\pm0.12$ & -2.80 $\pm0.10$ & 2.1 $\pm0.2$ \\
J175510.50-412812.1 & Gaia-ESO & 12.4 $\pm4.4$ & 5187 $\pm59$ & 2.23 $\pm0.31$ & -2.57 $\pm0.19$ & 2.0 $\pm0.2$ \\
 & This paper & 21.9 $\pm6.4$ & 5266 $\pm160$ & 1.75 $\pm0.13$ & -2.36 $\pm0.31$ & 2.3 $\pm0.2$ \\
J175652.43-413612.8 & Gaia-ESO & 5.4 $\pm2.8$ & 5035 $\pm196$ & 2.65 $\pm0.54$ & -2.48 $\pm0.23$ & 1.5 $\pm0.2$ \\
 & This paper & 3.1 $\pm0.9$ & 5142 $\pm160$ & 3.15 $\pm0.13$ & -2.39 $\pm0.13$ & 1.1 $\pm0.2$ \\
\hline
\end{tabular}
\end{minipage}
\end{table*}

In general, the two sets of parameters are very similar, and provide confidence in our method of determination. The average temperature offset (all offsets described are written as Gaia-ESO - new parameters) between the four stars is $-13\pm24$\,K, well within the predicted systematic uncertainties. Due to the wide variety of S/N between the four stars, and the differences in temperature determinations, it would perhaps be better to consider separately the two stars with high S/N, and the two with low S/N. In this case, the temperature offset for the "high-S/N" stars is $+68$\,K, but for the "low-S/N" stars $-93$\,K. The differences between the $\log{g}$ and [Fe/H] values for the two sets of analysis are also minor, well within the uncertainties. The average $\log{g}$ offset is $+0.03\pm0.10$\,dex. The average [Fe/H] offset is $-0.08\pm0.03$\,dex, and all four stars have individual differences of less than $0.25$\,dex. 

There are large differences in the uncertainties quoted for each analysis. The statistical uncertainties used in the GES analysis, whilst indicative of the size of the differences between the several nodes' analyses, do not accurately reflect the difference in uncertainty caused by the large variation in S/N of the spectra. In particular, the quoted uncertainties for SMSS J175510.50-412812.1 are smaller than for the other three stars, despite the low S/N of that star. 

These differences in parameters have led to offsets in the estimated distances to the stars, particularly for the low-S/N pair. The average offset is $-2.0$\,kpc, the average for the high-S/N stars is only $-0.4$\,kpc, well within the rather large distance uncertainties. Both low-S/N stars have very different distances - altering their position in the Galaxy from both being consistent with being in the bulge (as shown in Figure 4 of \citet{2014MNRAS.445.4241H}) to most likely not being in the bulge.
\begin{figure}
  \centering
  \includegraphics[width=0.99\columnwidth]{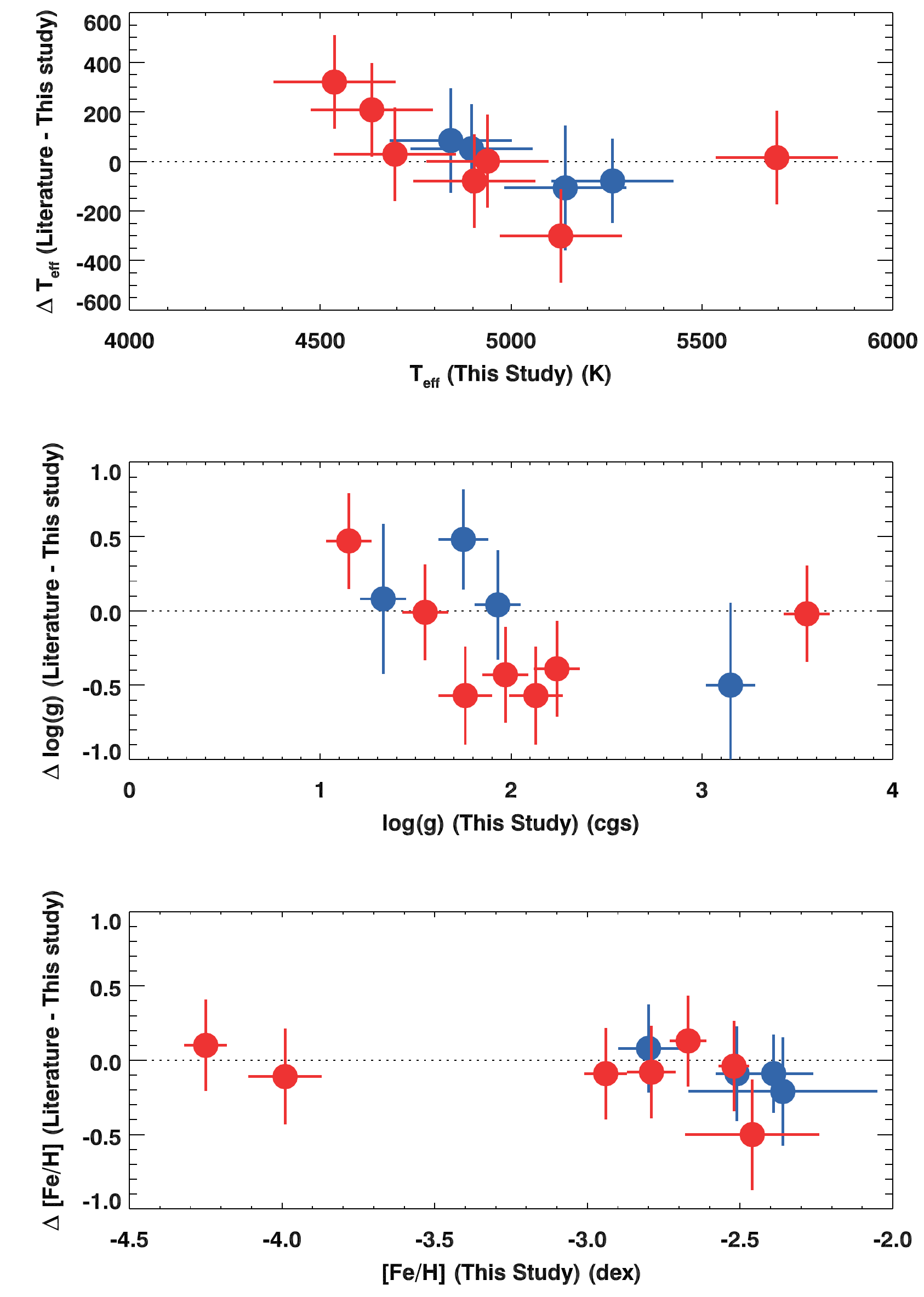}
  \caption{Comparison of the stellar parameters of the four GES stars (blue dots), along with seven halo stars from the literature (red dots). The parameters are derived in this paper, compared with those from \citet{2014MNRAS.445.4241H}, and the literature values \citep{2013ApJ...762...26Y}.}
  \label{fig:complit}
\end{figure}

To further investigate the reliability of our methodology for determining stellar parameters, we have also derived parameters for seven literature stars. We took five halo stars from \citet{2013ApJ...762...26Y}, where the spectra were also observed using MIKE on Magellan\footnote{The stars used are CD --38$\degr$ 245, CS 22892--052, CS 30336--049, HE 2142--5656, and HE 2247--7400.}. Furthermore we analysed the two Gaia benchmark metal-poor stars HD122563 and HD140283, using the publicly available UVES-POP spectra \citep{2003Msngr.114...10B}, but have taken the parameters for these stars from \citet{2013ApJ...762...26Y}, for consistency. The results of this comparison are shown in Figure \ref{fig:complit}, which reveals no obvious trends in the offsets between the two parameter sets, with our [Fe/H] measurements in very good agreement with the literature values. The mean offset for [Fe/H] is $-0.08$\,dex, and the standard error of the mean, $0.03$\,dex, which is well within the levels of the reported uncertainties. For $\log{g}$ the average difference is $-0.22$, with a standard error of the mean of $0.06$. The differences in $T_{\rm{eff}}$ are more noticeable, particularly at low temperatures. Of the two stars with the lowest $T_{\rm{eff}}$ values in our analysis, one is HD122563. The Balmer line temperature we find for that star is $4635$\,K, whereas the value in \citet{2013ApJ...762...26Y} is $4843$\,K (from photometry). Other literature analyses for this star give temperatures of $4665$\,K (\citealt{2012MNRAS.427...27B}, a non-LTE analysis) and 4587\,K (\citealt{2015A&A...582A..49H}, temperatures taken from interferometry), which are much closer to our value. The coolest star, CD --38$\degr$ 245 ([Fe/H]$=$$-4.0$), has a difference of $300$\,K between our two methods, but it is not clear why. On average, however, there is only a mean separation between our temperatures of $+28$\,K, with a standard error of $28$\,K. We thus conclude that our stellar parameters are on a scale consistent with other state-of-the-art analyses of metal-poor stars. 

\begin{table*}
\begin{minipage}{180mm}
\centering
\caption{Chemical abundances for all 14 stars. The full version of the table can be found online.}
\label{table:abunds}
\begin{tabular}{lrrrrrrrr}
\hline
Star (SMSS) & [C/Fe] & [O/Fe] & [Na/Fe] &[Mg/Fe] & [Al/Fe] & [Si/Fe] & [K/Fe] & ... \\
\hline
J182637.10-342924.2 & 0.68 $\pm0.20$ & - 				& -0.36 $\pm0.19$ & 0.50 $\pm0.17$ & -0.93 $\pm0.26$ & 0.26 $\pm0.13$ & 0.76 $\pm0.19$ & ...\\
J182600.09-332531.0 & -0.41 $\pm0.22$ & - 			& -0.10 $\pm0.25$ & 0.16 $\pm0.14$ & -0.3 $\pm0.25$ & 0.60 $\pm0.16$ & 0.48 $\pm0.17$ & ...\\
J182601.24-332358.3 & 0.32 $\pm0.20$ & - 				& 0.27 $\pm0.21$ & 0.78 $\pm0.13$ & -0.45 $\pm0.19$ & 0.56 $\pm0.22$ & 0.69 $\pm0.16$ & ...\\
J182753.81-334607.7 & -0.41 $\pm0.25$ & -				 & -0.40 $\pm0.18$ & 0.29 $\pm0.15$ & -1.42 $\pm0.22$ & 0.34 $\pm0.23$ & 0.55 $\pm0.19$ & ...\\
... & ...& ... & ...& ... & ... & ... & ... & ... \\
\hline
\end{tabular}
\end{minipage}
\end{table*}

\section{Results and Discussion}

\subsection{Positions and Kinematics}

\begin{figure}
  \centering
  \includegraphics[width=0.99\columnwidth]{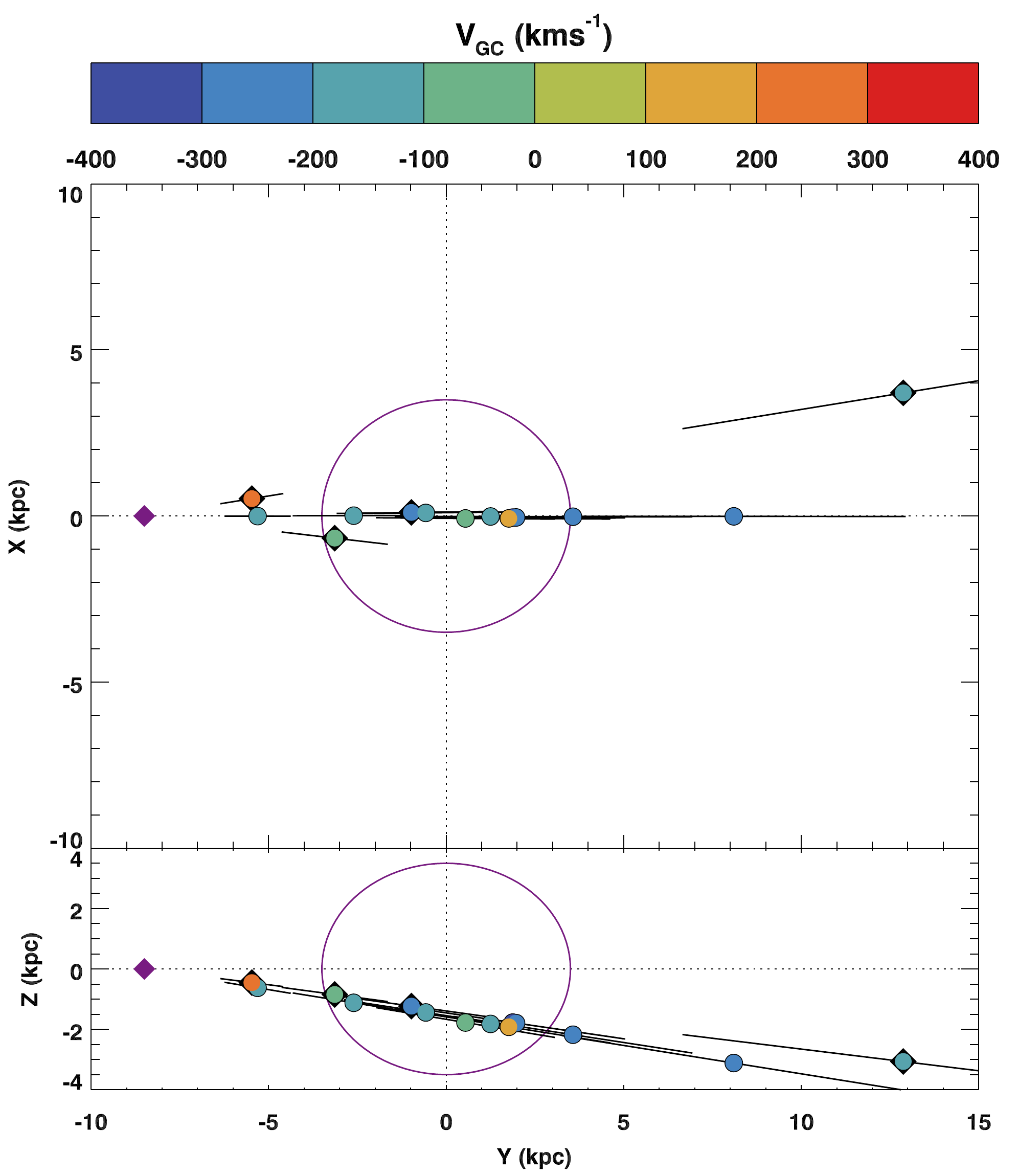}
  \caption{A diagram showing the positions of our stars in the Milky Way, viewed from above and from the side, with the Galactic centre at ($0, 0$) and the Sun at ($-8.5, 0, 0$), shown as the purple diamond. The purple circle represents a 3.43\,kpc radius sphere around the centre of the Galaxy. Each star is coloured according to its Galactocentric velocity, with the error bars calculated in the projection plane from the distance uncertainties. The GES stars are shown as diamond symbols.}
  \label{fig:positions}
\end{figure}

All the stars were originally chosen from the SkyMapper bulge fields shown in Fig. \ref{fig:pos}, covering roughly a 20$\degr$x20$\degr$ area centred on the Galactic centre. The stars analysed here were chosen after the first observing runs in 2012, which is why the majority are all from one field, close to ($l, b$)=(0, $-10$). Combining their positions with the derived distances, we can confirm that the majority of the stars in the sample are situated within the Milky Way bulge. Figure \ref{fig:positions} shows the positions of each of the stars as viewed from above and from the side of the Galaxy, with the Sun positioned at ($-8.5, 0, 0$) \citep{2012ApJ...759..131B}. A circle of radius 3.43\,kpc demonstrates the simplest criteria on bulge membership \citep{2012A&A...538A.106R}, although we note here that the bulge is actually a more complex bar shape, extending further out into the plane along the Y axis, and reaching above the plane into a "peanut" shape, as seen in recent work matching models to bulge data \citep{2010ApJ...720L..72S,2013MNRAS.435.1874W,2014ApJ...787L..19N,2015MNRAS.447.1535N}. Nine out of the 14 stars analysed here lie within this 3.43\,kpc radius, and another three lie just outside the circle (two in the foreground, one in the background), which, when considering the stellar parameters uncertainties and a more realistic bulge, could well be members also. The distance uncertainty for the two stars seemingly located furthest away are so large that they are also consistent with residing in the bulge.

A key component of the assertion that metal-poor stars in the bulge are truly among the oldest stars surviving in the Milky Way is the question of whether these stars are "true" bulge stars, or merely halo stars just passing through the bulge region on eccentric orbits. Currently, this is a question that cannot be easily answered due to the lack of published, reliable proper motions and thus meaningful orbit information of these metal-poor stars. There are proper motions for these 14 stars in existing catalogues, but the large uncertainties of approximately 5\,mas\,yr$^{-1}$ mean that any orbits derived are very uncertain and open to the possibility of being either bulge or halo orbits. Until the publication of new proper motion catalogues of the outer bulge, such as OGLE-IV \citep{2015AcA....65....1U} or VVV \citep{2010NewA...15..433M}, we cannot eliminate either possibility that these stars have bulge-like or halo-like orbits. Accurate distances would also improve the levels of uncertainties for the orbits; here parallaxes from Gaia should be helpful in the near future.

A potentially good, independent tracer of the relevant stellar population are RR Lyrae, as they are old, metal-poor, and have standardisable distances. The BRAVA-RR survey of Galactic bulge RR Lyrae stars has found that the fraction of RR Lyrae stars toward the bulge with orbits more consistent with a halo-rather-than-bulge dynamical origin may be as low as 1\% \citep{2015arXiv150602664K}.
\begin{figure}
  \centering
  \includegraphics[width=0.8\columnwidth]{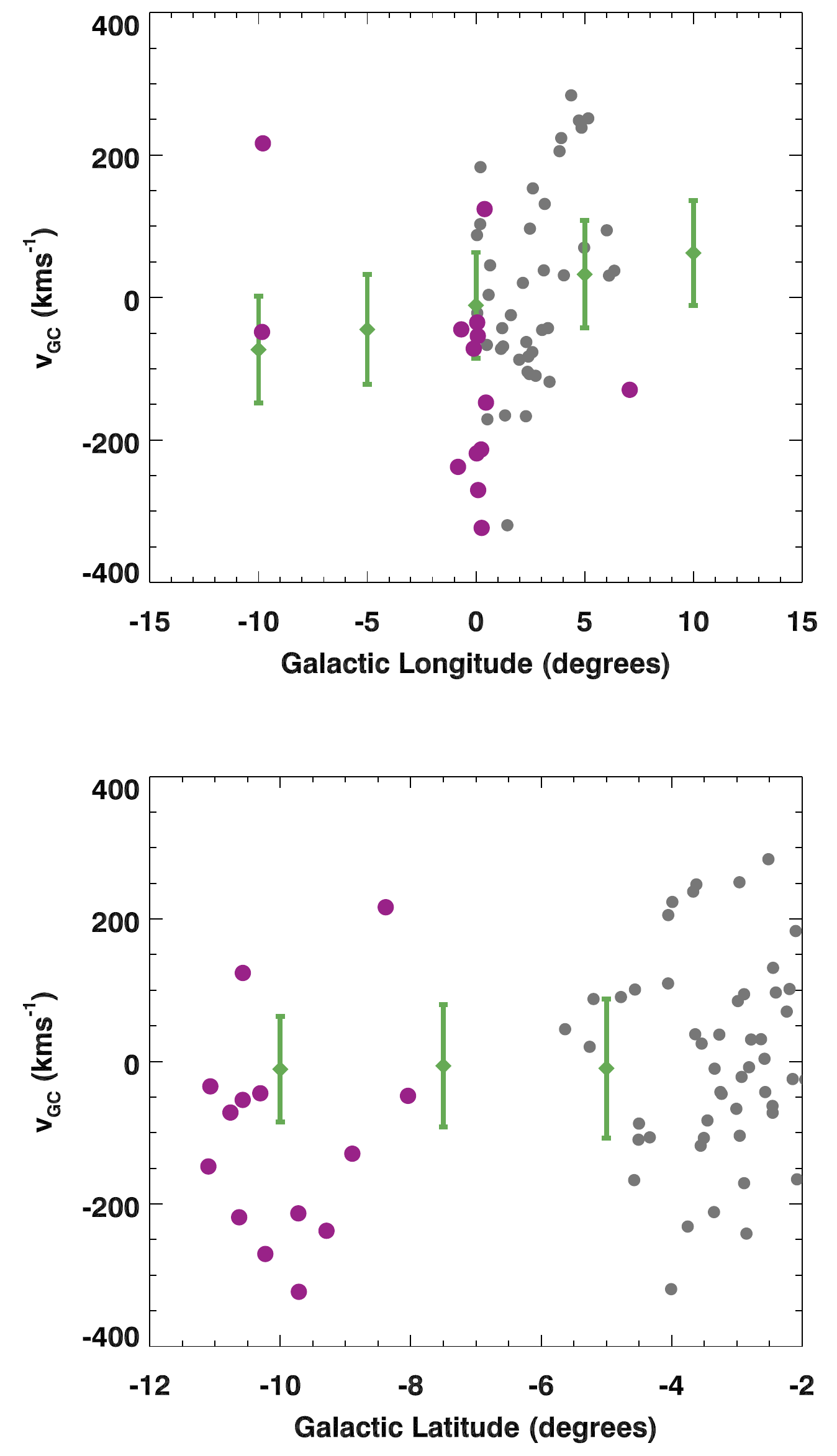}
  \caption{Plots comparing the Galactocentric velocity to both Galactic longitude and latitude. Our sample is shown as purple circles, compared with the bulge dwarf stars of \citet{2013A&A...549A.147B} (grey circles), and the velocity dispersions of certain ARGOS fields \citep{2013MNRAS.432.2092N} at similar latitudes and longitudes (green lines).}
  \label{fig:v_gc}
\end{figure}

We can consider the radial velocities of the stars - which we have determined from their spectra to an accuracy of $\sim$$1$\,km\,s$^{-1}$ - to help constrain their possible dynamics. To compare the velocities to other samples of bulge stars, they have been converted into Galactocentric velocities, which correct for the solar motion around the Galaxy. To do this we have used the equation originally from \citet{1981gask.book.....M}, given in \citet{2013MNRAS.432.2092N}:
\begin{equation}
\label{eq:gal}
\begin{aligned}
V_{GC} = &V_{helio} + 220\sin{l}\cos{b} \\
&+ 16.5(\sin{b}\sin{25}+\cos{b}\cos{25}\cos{[l-53]}),
\end{aligned}
\end{equation}
where $V_{helio}$ is the star's heliocentric velocity, and $l$ and $b$ are the Galactic coordinates in degrees.
The velocity dispersion amongst these 14 stars is quite large -- $\sigma=149.6$\,km\,s$^{-1}$ -- compared to the velocity dispersion found by the ARGOS survey \citep{2013MNRAS.432.2092N} in the closest field to our sample at ($l, b$) = (0, $-10$) of $\sigma=74.3$\,km\,s$^{-1}$. Whilst the velocity dispersion of the EMBLA stars is more characteristic of a halo population, the small number of stars involved makes it difficult to say anything meaningful about this quantity. Instead, we have compared our sample of stars to the microlensing sample of \citet{2013A&A...549A.147B} (Figure \ref{fig:v_gc}). It is clear that both samples have a wide range in velocities, spanning more than 700\,km\,s$^{-1}$, and that for a small sample, the EMBLA stars do not appear more dispersed than the \citet{2013A&A...549A.147B} stars. There is a positional offset between the two groups; the \citet{2013A&A...549A.147B} stars are found closer to the Galactic plane, $b\ge$$-6$, whereas the EMBLA stars are at high negative latitudes, $b<$$-8$. The ARGOS sample spans both of these positions, and finds no significant difference in the velocity dispersion between the two, so it is unlikely that this offset could make a difference to what we expect dynamically of the two sets.

\subsection{Stellar Parameters}
\begin{figure}
  \centering
  \includegraphics[width=0.99\columnwidth]{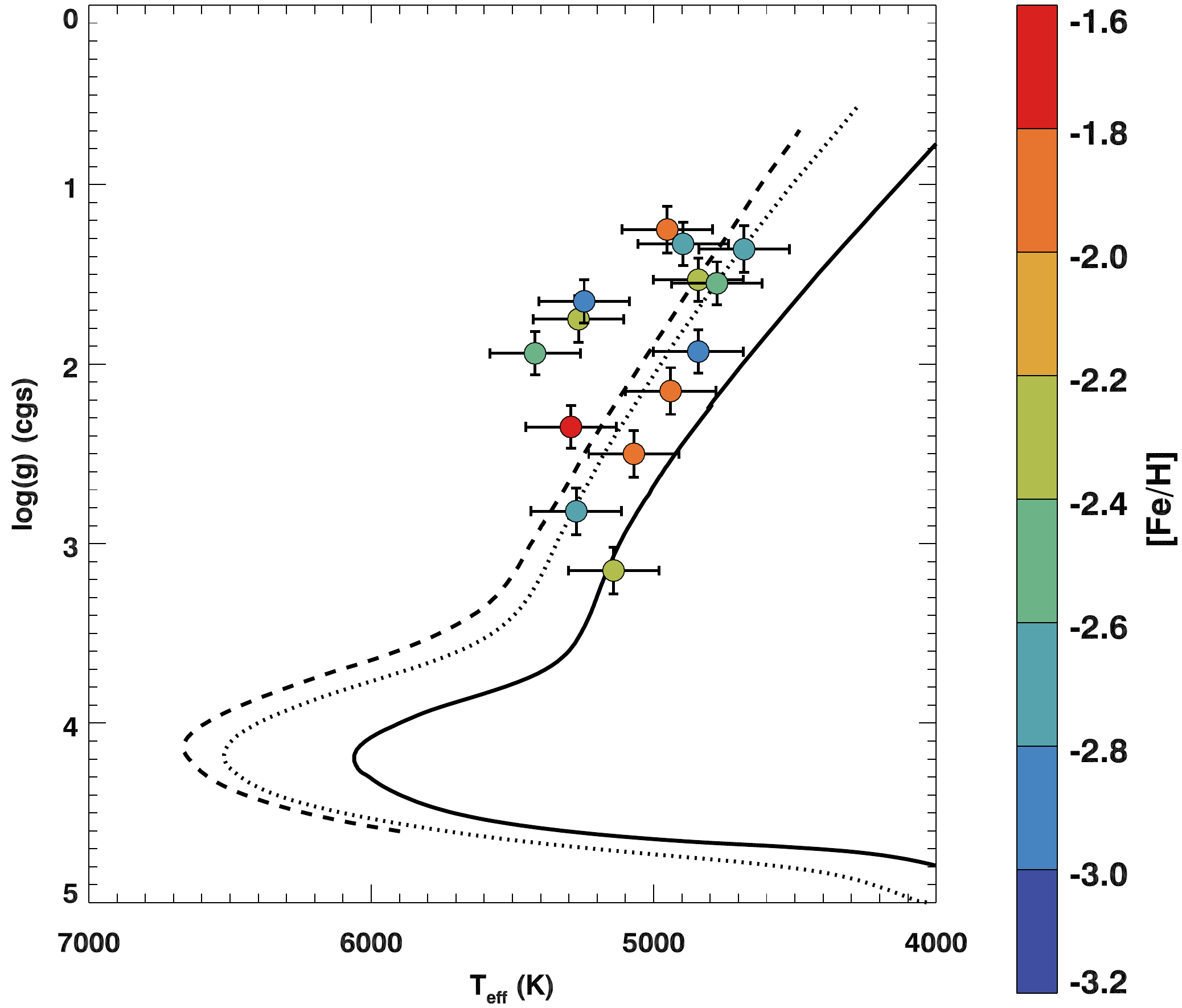}
  \caption{The HR diagram for all 14 stars discussed, given different colours to indicate their metallicity. Also shown are three 14\,Gyr Dartmouth isochrones \citep{2008ApJS..178...89D}, at metallicities of [Fe/H]$=-1$ (solid line), $-2$ (dotted line), and $-3$ (dashed line). All three are $\alpha$-enhanced, with [$\alpha$/Fe]$=0.4$.}
  \label{fig:hrdiagram}
\end{figure}

As part of the photometric selection for EMBLA, we attempted to limit the stars to those on the RGB using the colour-magnitude diagram (CMD). However, those stars observed in high-resolution here were all selected from early fields observed at the AAT during our pilot observations. At that time, the CMD cut had not been introduced into our selection process. Therefore the numbers of dwarfs that were observed is much higher than in fields observed later. Despite this, the stars chosen for high-resolution follow-up were confirmed spectroscopically as giants (as displayed on the HR diagram in Figure \ref{fig:hrdiagram}), and as mentioned previously, mostly contained within the bulge region. 

The metallicities of the stars in our sample are some of the lowest found in the Galactic bulge. \citet{2014ApJ...797...13S} used IR photometry to find bright metal-poor stars in the bulge, and have discovered three stars with metallicities of [Fe/H]$=$$-3.02$, $-2.84$, and $-2.70$, similar in metallicity to the most metal-poor stars here. Between the EMBLA results (\citealt{2014MNRAS.445.4241H,2015Natur.527..484H}), the APOGEE metal-poor bulge stars \citep{2013ApJ...767L...9G}, and the \citet{2014ApJ...797...13S} stars, a total of 37 stars have now been studied in high-resolution with [Fe/H]$<$$-2.0$, providing the observational evidence that metal-poor stars do exist in the central regions of the Galaxy. Now that a significant number of these have been found, the need for further detailed work into these stars' kinematics is vital in order to distinguish which have the lowest binding energies, as these would be more likely to have formed at redshifts $z>15$ \citep{2010ApJ...708.1398T}.

\subsection{Chemical Abundances}
\begin{figure*}
\begin{minipage}{180mm}
  \centering
  \includegraphics[width=0.99\columnwidth]{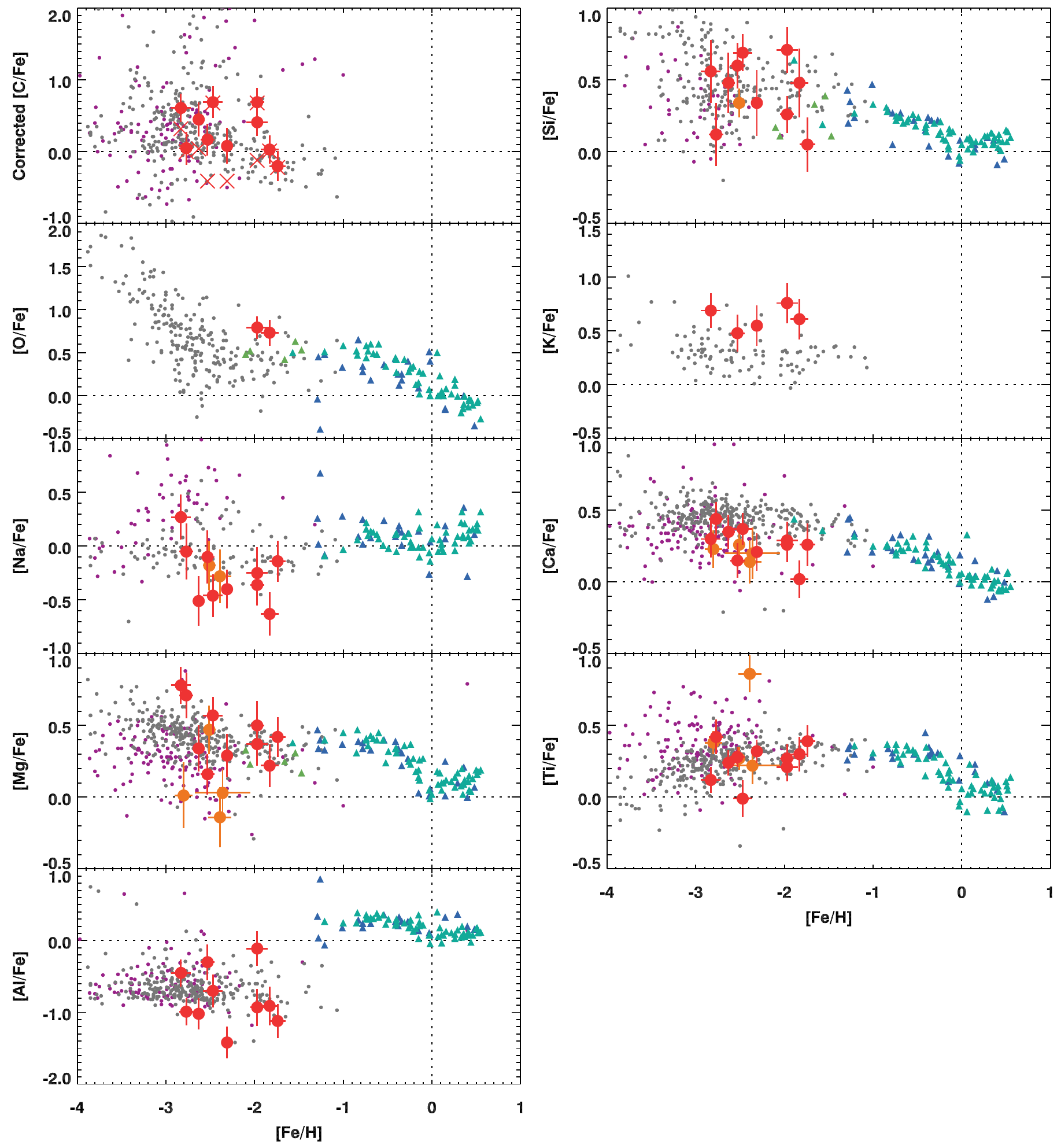}
  \caption{Abundance trends of elements C to Ti, with respect to [Fe/H]. Both the Magellan data (red circles) and Gaia-ESO data (orange circles) are shown.  For comparison, we show the literature samples of both bulge (triangles) and halo (dots) stars, taken from \citealt{2013A&A...549A.147B} (turquoise), \citealt{2010A&A...513A..35A} (blue), \citealt{2013ApJ...767L...9G} (green), \citealt{2014AJ....147..136R} (grey), and \citealt{2013ApJ...762...26Y} (purple). The original observed C values are shown as red crosses in the C plot. }
  \label{fig:abunds1}
\end{minipage}
\end{figure*}

\begin{figure*}
\begin{minipage}{180mm}
  \centering
  \includegraphics[width=0.99\columnwidth]{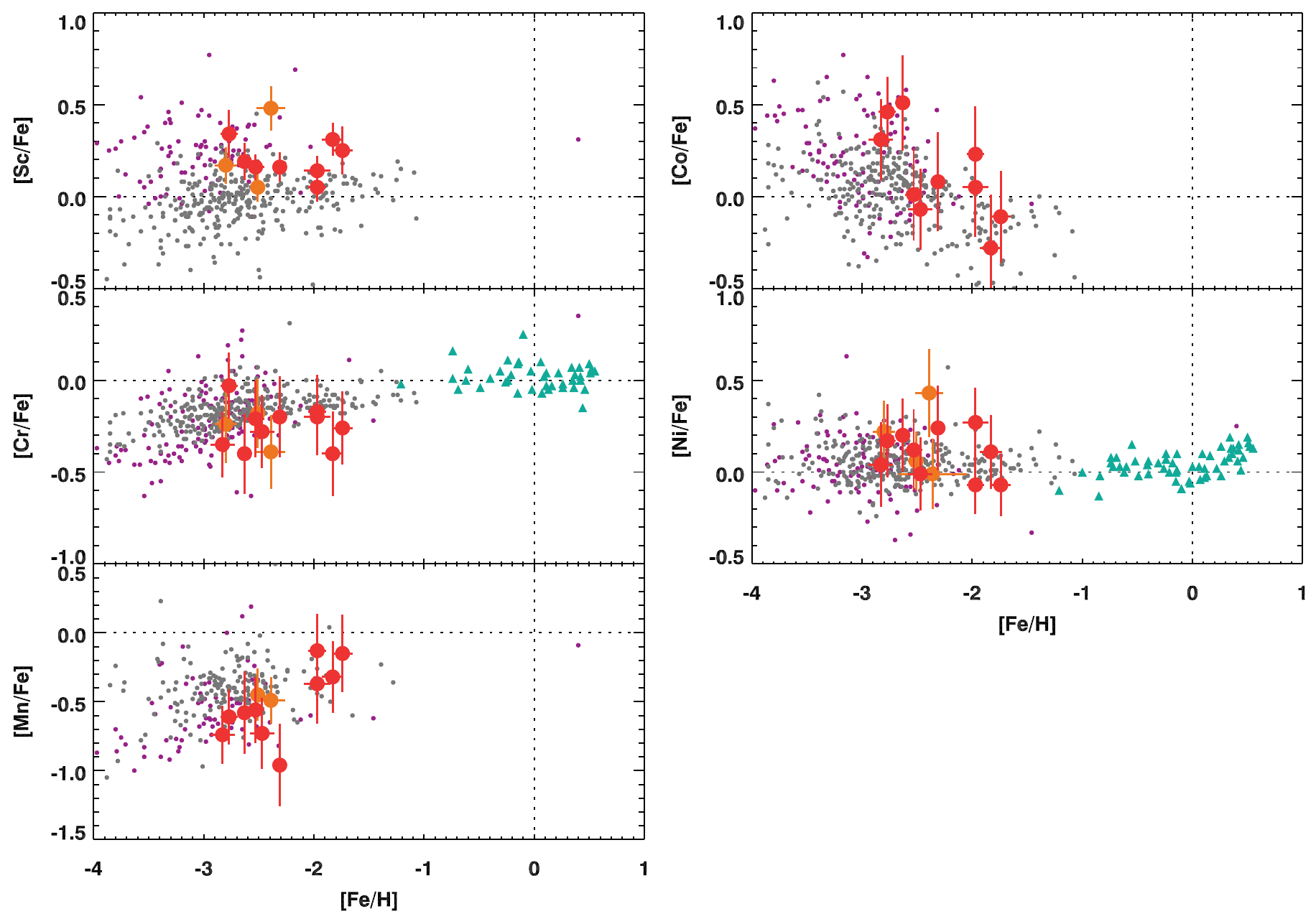}
  \caption{Abundance trends for the iron-peak elements, symbols as in Figure \ref{fig:abunds1}. }
  \label{fig:abunds2}
\end{minipage}
\end{figure*}

\begin{figure*}
\begin{minipage}{180mm}
  \centering
  \includegraphics[width=0.99\columnwidth]{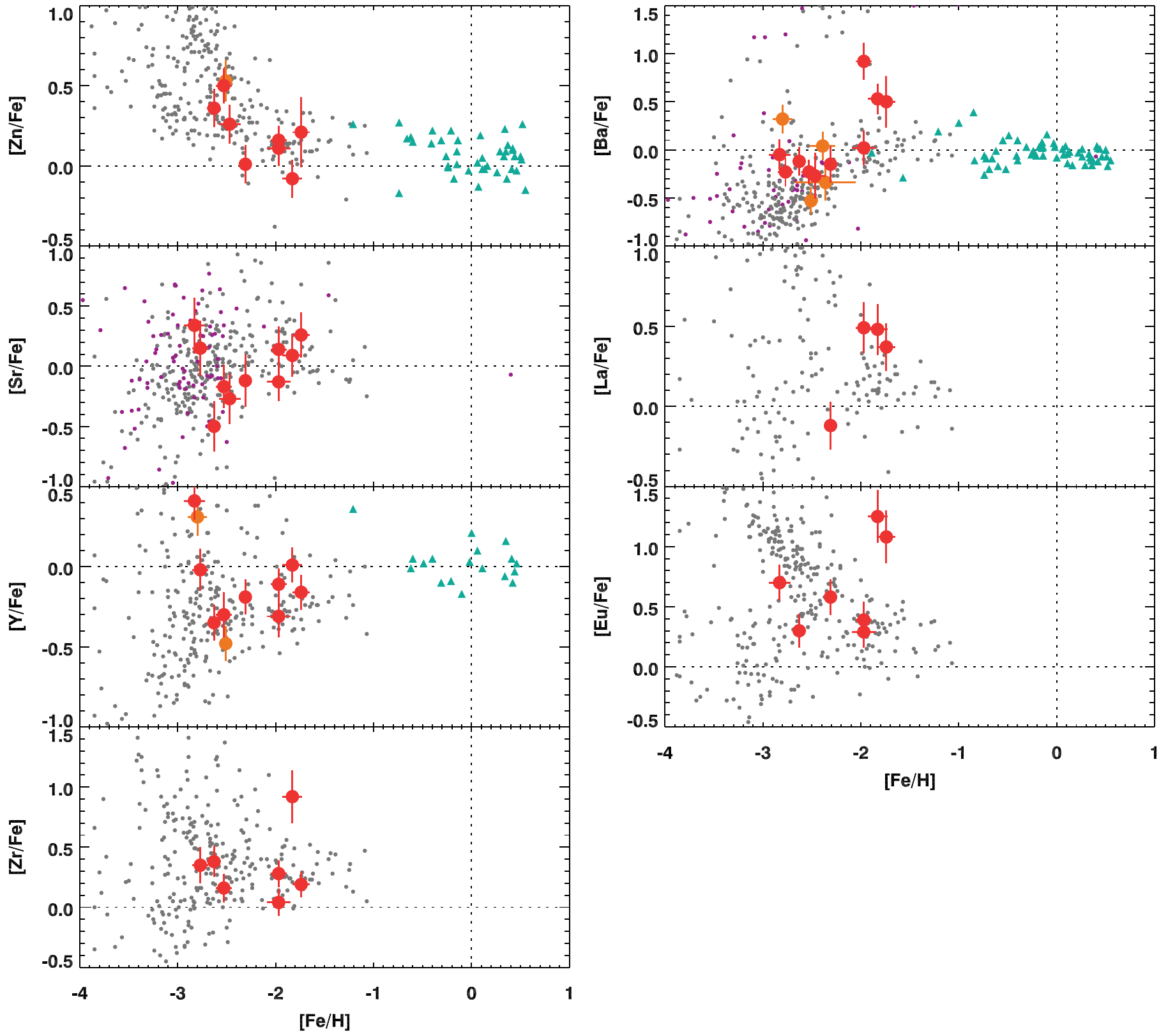}
  \caption{Abundance trends for the neutron capture elements, symbols as in Figure \ref{fig:abunds1}. }
  \label{fig:abunds3}
\end{minipage}
\end{figure*}

The elemental abundances that have been measured are displayed in Figures \ref{fig:abunds1}, \ref{fig:abunds2}, and \ref{fig:abunds3}, shown with respect to their [Fe/H] values. Also shown are bulge and halo samples taken from the literature \citep{2013A&A...549A.147B,2010A&A...513A..35A,2013ApJ...767L...9G,2014AJ....147..136R,2013ApJ...762...26Y}. In case there are any noticeable differences between the sample observed as part of GES or the sample observed on MIKE, perhaps caused by systematic differences between the two methods of observation, they are shown as different colours - however, there does not appear to be any significant difference.

\subsubsection{Carbon}
The carbon abundances in the sample have a wide range, covering values from as low as [C/Fe]$=$$-0.41$, up to $0.68$. This wide scatter is typical of metal-poor stars, like those found in the halo. Interestingly, however, is the lack of stars with much higher C abundances. For some time now, it has been accepted that metal-poor stars broadly fit into two categories; the C-normal stars and the C-enhanced (CEMP) stars (first categorised in \citet{2005ARA&A..43..531B}; \citet{2013ApJ...762...28N} present a recent discussion on the differences between the two populations). Whether one considers the criterion for a carbon-enhanced star to be [C/Fe]$>1.0$ \citep{2005ARA&A..43..531B} or [C/Fe]$>0.7$ \citep{2007ApJ...655..492A}, none of these bulge stars make the grade. There are several reasons this could be, firstly, as the stars have already started to evolve up the red giant branch, their photospheric C abundances may have become depleted due to envelope mixing. \citealt{2014ApJ...797...21P} have calculated the corrections needed to find the original carbon abundances of evolved stars, and tabulated these corrections in terms of $\log{g}$ and [Fe/H]. We have applied these theoretical corrections to our stars and displayed the result in Figure \ref{fig:abunds1}, which shows the C levels the stars would have been born with, rather than that observed now. Whilst these corrections have decreased the number of stars with sub-solar C levels, the stars with the higher C abundances have been largely unaffected, as their surface gravities imply they are less evolved. Even with the correction, none of the bulge stars have [C/Fe]$\ge0.7$.

A second reason for a lack of CEMP stars in our sample could be a selection bias introduced by our SkyMapper photometric selection. A recent paper on the SkyMapper halo metal-poor search \citep{2015ApJ...807..171J} has found no CEMP stars with [C/Fe]$\ge2$, and suggested that, due to the strong CH absorption of such stars, photometric colours may have been affected, placing them outside the selected region in the colour-colour diagram (Fig. \ref{fig:photometry}). \citet{2015ApJ...807..171J} do not describe a lack of CEMP stars, however, finding 20\% of their sample have [C/Fe]$\ge0.7$, in line with the values expected from previous studies. Therefore, whilst possible that the EMBLA Survey could have missed stars with extremely large C enhancements due to selection effects, stars with mild C enhancements should still have been found.

Perhaps the most likely explanation for the lack of CEMP stars is small number statistics. This sample contains 10 stars with measured C abundances; the literature suggests that 21\% of stars with [Fe/H]$<$$-2.0$ should have [C/Fe]$\ge$$1.0$ \citep{2006ApJ...652L..37L}, so we would only expect to find approximately two CEMP stars. Only with a larger sample can we confirm that there are no, or very few, CEMP stars in the bulge. If this is confirmed, it would have wide ranging implications for the studies of first star formation. Overabundances of C have been frequently cited as the observational evidence that C and O are required to provide the cooling mechanisms needed for early gas clouds to condense into stars at the very lowest iron abundances \citep{2007MNRAS.380L..40F}. \citet{2010ApJ...708.1398T} suggested that the number of CEMP stars should increase with increasing age of the population studied - indicating that the CEMP fraction should be larger in the bulge than the halo - the opposite of our current results, while mindful of the relatively small number statistics.  

\subsubsection{Light Elements - O, Na, Al, K}
We measured O in those stars where the forbidden line at 630.0\,nm was detectable, but unfortunately due to the metal-poor nature of the stars, this was only possible for two of them, both with [Fe/H]$\approx -2.0$. The two measurements we have are unusually high - averaging $0.76$ whereas the APOGEE sample \citep{2013ApJ...767L...9G} at approximately the same metallicity average $0.52$ across five stars. This difference could be due to the different lines measured; the APOGEE spectra are in the infrared with the  oxygen abundances measured from OH lines. 

The Na abundances have been corrected for non-LTE effects using \citet{2011A&A...528A.103L}, as have the abundances for the stars in the \citet{2014AJ....147..136R} sample (grey circles), and on the whole, our abundances match well, perhaps our Na abundances are slightly lower. A similar conclusion can be drawn for the Al abundances, which, like Na, are predominately sub-solar. Non-LTE effects have not been considered for our stars nor the literature sample, it is expected that these corrections would bring the abundances closer to the solar [Al/Fe] value \citep{2008A&A...481..481A}. It appears that the dispersion in our Al abundances is larger than the \citet{2014AJ....147..136R} sample, similar to that found in the alpha elements. Our K abundances are noticeably higher than those in the halo literature \citep{2014AJ....147..136R}, although ours have not been corrected for non-LTE effects, whereas the literature stars have (using \citealt{2002PASJ...54..275T}) and it was noted that the corrections resulted in lower abundances than before.

\subsubsection{Alpha Elements}
One of the most intriguing results found in \citet{2014MNRAS.445.4241H} was the intrinsic scatter in the $\alpha$-elements, particularly in Mg and Ti, a marked difference from literature metal-poor halo stars. This is further supported with the larger dataset presented here, although we caution again that the sample is too small to draw final conclusions. It is noticeable that Si is also affected - the dispersion of Mg, Si and Ti are respectively $\sigma=0.26$, $0.22$, and $0.19$. The average measurement uncertainties for these elements are $0.13$, $0.14$, and $0.10$; the dispersion cannot easily be explained by the size of the uncertainties. Compared to halo stars of the same metallicity the abundances found in the EMBLA sample are more dispersed: the spread of Mg in the \citet{2013ApJ...762...26Y} giant sample about a linear fit is $\sigma=0.13$, and for those stars with [Fe/H]$>-3.0$ in \citet{2014AJ....147..136R}, the dispersion of Mg is $\sigma=0.12$. It should be noted that the offsets between the \citet{2013ApJ...762...26Y} sample and the \citet{2014AJ....147..136R} sample for Mg and Ti can be explained by systematic differences in their analyses, as discussed in Section 9.4 of \citet{2014AJ....147..136R}. 

There are several examples of bulge stars with extreme differences between the different elements, for example SMSS J182948.48-341053.9, which has [Si/Fe]$=0.69$ and [Ti/Fe]$=-0.01$, and SMSS J175652.43-413612.8 with [Mg/Fe]$=$$-0.18$ and [Ti/Fe]$=$$0.86$. Only a couple of stars with large variations in their $\alpha$ abundances have been found in the halo, such as HE 2136-6030 \citep{2013ApJ...762...26Y} with [Si/Fe]$=$$1.20$ but [Mg/Fe]$=$$0.08$. These large differences perhaps suggest inhomogeneous mixing at the time of formation, indicative that these stars were formed early in the life of the Universe. 

\begin{figure}
  \centering
  \includegraphics[width=0.99\columnwidth]{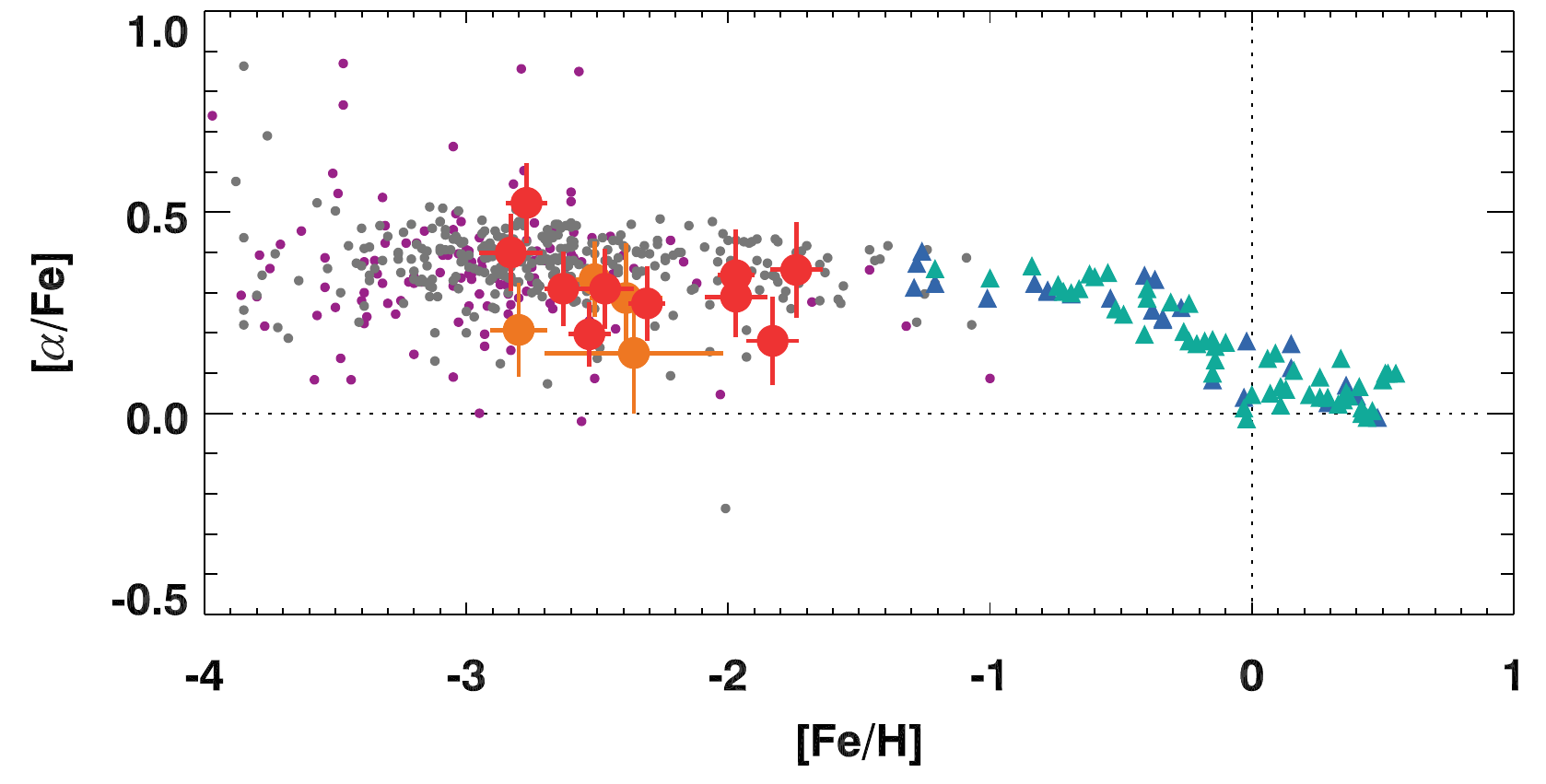}
  \caption{The $\alpha$ abundance trends of the 14 stars, compared to literature values (symbols as in Fig. \ref{fig:abunds1}). Here [$\alpha$/Fe] = ([Mg/Fe]+[Ca/Fe]+[Ti/Fe])/3.}
  \label{fig:alpha}
\end{figure}

The caveat to this story of a wide scatter between different $\alpha$-elements and different stars is Ca. The fourteen stars show a much tighter trend in [Ca/Fe], with a dispersion of only $\sigma=0.11$, and they appear to plateau in a similar way to halo stars and stars from the thick disc. 
Averaging over the alpha elements (Figure \ref{fig:alpha}), we find they are at the level we would expect for metal-poor stars - that is, $\alpha$-enhanced, suggesting the gas they were formed from was polluted by fairly massive core-collapse supernovae. It is interesting to note that, for three of the four $\alpha$-elements, the level of enhancement is slightly lower than that seen in the halo (noticeable in [Ca/Fe] when compared to the \citet{2014AJ....147..136R} sample), which was similarly noted in the previous study of metal-poor bulge stars by \citet{2013ApJ...767L...9G}. The average values and uncertainties of the four compared to the giants of the \citet{2013ApJ...762...26Y} sample in parentheses are [Mg/Fe]$=0.35\pm0.13$ (0.30), [Si/Fe]$=0.42\pm0.14$ (0.57), [Ca/Fe]$=0.25\pm0.10$ (0.32), and [Ti/Fe]$=0.31\pm0.10$ (0.32) (bearing in mind that we correct both Mg and Ca for non-LTE effects). A larger sample of stars is needed to disentangle any trend from the measurement uncertainties.

\subsubsection{Iron-peak Elements}
All iron-peak elements measured in this work match well when compared to the literature values in the halo. It is expected that both Cr and Mn have large positive NLTE corrections \citep{2010A&A...522A...9B}, which, when applied would bring them close to the solar values (the halo literature stars have also been measured with LTE assumptions, so would have similar NLTE corrections). Unlike in \citet{2015ApJ...809..110C}, no unusually low abundances of Sc or Mn are found in these stars.

\subsubsection{Neutron-capture Elements}
In general, the abundances of the neutron-capture elements measured here match the trends found in halo metal-poor stars. The overriding feature of these elements in halo stars is a very large scatter; this has been found in all studies of metal-poor stars to date \citep{1995AJ....109.2757M, 2007A&A...476..935F,2013ApJ...762...26Y, 2014AJ....147..136R}, and there have been numerous attempts to explain the various abundance patterns seen (e.g., \citealt{2004ApJ...601..864T}, \citealt{2010ApJ...724..975R}). One such explanation is the introduction of yields from fast-rotating massive stars (or spinstars), which would produce elements formed through the s-process \citep{2012RvMP...84...25M}. Typically the s-process occurs in AGB stars, which are less massive than the stars that undergo core-collapse supernovae, and so it occurs on longer timescales \citep{1999ARA&A..37..239B}. Therefore fast-rotating massive stars provide a theoretical reason why s-process elements are found in metal-poor stars; works such as \citet{Cescutti:2013hl} and \citet{2014A&A...565A..51C} have explained the trends seen in the halo.

\begin{figure}
  \centering
  \includegraphics[width=0.99\columnwidth]{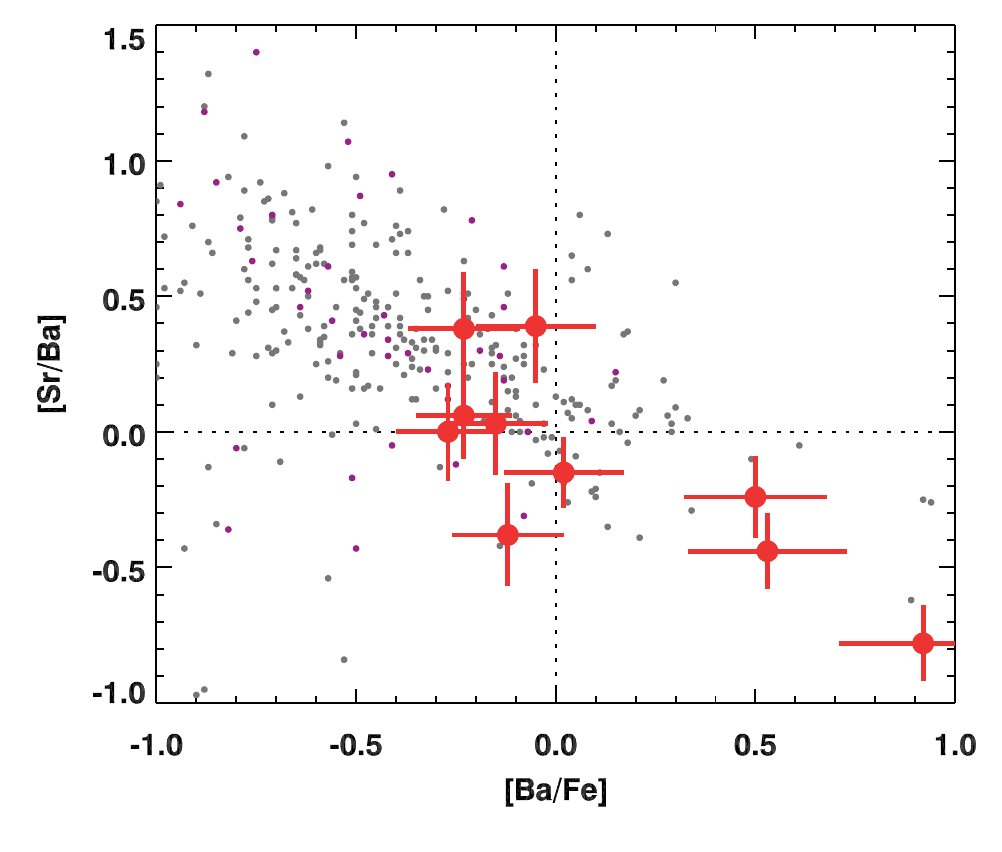}
  \caption{[Sr/Ba] with respect to [Ba/Fe], compared to literature values (symbols as in Fig. \ref{fig:abunds1}).}
  \label{fig:srba}
\end{figure}

Previous studies of bulge stars have also attempted to find the signatures of spinstars \citep{2014A&A...570A..76B} at higher metallicities than those studied here. \citet{Cescutti:2013hl} show that incorporating spinstar models into the chemical evolution model leads to a larger scatter of $\alpha$-elements, as we have found. However, one of the key indicators of these spinstars is a high [Sr/Ba] ratio, ($>0.5$), which is not borne out in our data (Figure \ref{fig:srba}). 

\begin{figure}
  \centering
  \includegraphics[width=0.99\columnwidth]{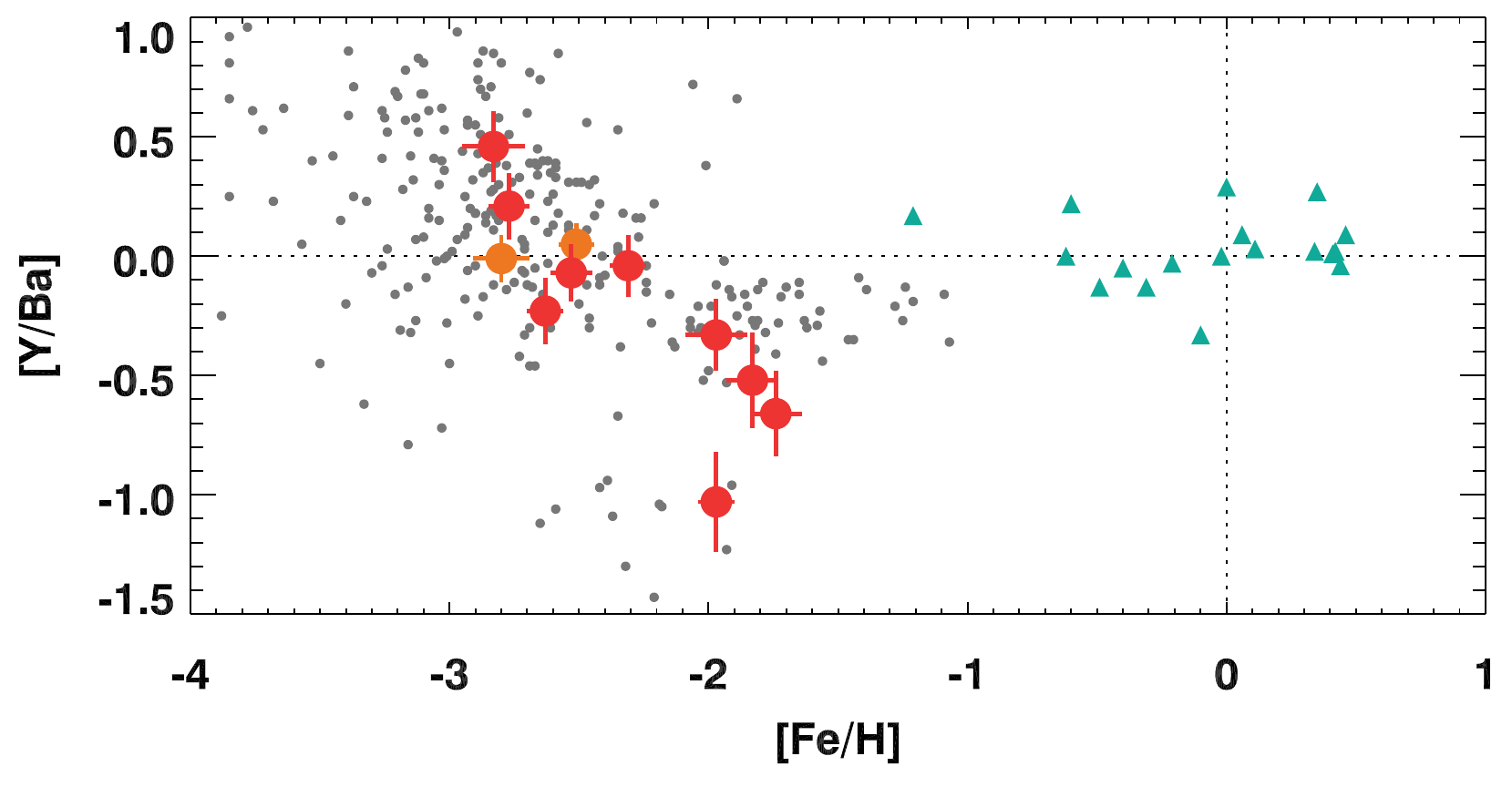}
  \caption{[Y/Ba] with respect to [Fe/H], compared to literature values (symbols as in Fig. \ref{fig:abunds1}).}
  \label{fig:yba}
\end{figure}

Another key indicator is high Y abundances, particularly when compared to Ba, and again our data show the opposite of this (Fig. \ref{fig:yba}). Our bulge stars have relatively low abundances of both Sr and Y. The Ba abundances too are predominately sub-solar, but there is one star (SMSS J182637.10-342924.2) with a high Ba abundance ([Ba/Fe]$=$$0.92$). This star has a very low [Sr/Ba] ratio, more so than any of the rest of the sample, and of many of the literature comparison stars. Studies into such "low-Sr/Ba" stars have suggested two explanations \citep{2014A&A...571A..40S}; stars in a binary system where the other star has evolved past the AGB phase and polluted the stellar atmosphere through mass transfer, or stars which themselves have started to undergo the mixing involved in the AGB. The parameters of the star (T$_{\rm{eff}}=5070$\,K, $\log{g}=2.50$) would make it highly unlikely that the star is an AGB star (Fig. \ref{fig:hrdiagram}), making the binary hypothesis more likely. Interestingly the star's C abundance is the highest of all the stars in the sample ([C/Fe]=0.68), close to the limit categorising it as a CEMP star. The star has a high ratio of $\log \epsilon$ (La/Eu)$=0.68$, indicative of s-process enriched material \citep{2008ARA&A..46..241S}, further suggesting that the star has received AGB pollution from a companion. Studies into possible radial velocity variations in the star would be needed to confirm the binarity. 


The $\log \epsilon$ (La/Eu) ratio is used as a strong indicator of s- and r- process enrichment, as shown in Figure 12a of \citet{2008ARA&A..46..241S}. Asides from SMSS J182637.10-342924.2, we have only been able to measure La in three other stars, as the lines are too weak at these metallicities. All three stars have very low $\log \epsilon$ (La/Eu) values, $-0.12$, $-0.13$, and $-0.19$. Not only is this value well below that expected for s-process enrichment, it is also below the predicted values for r-process enrichment. \citet{2010ApJ...724..975R} calculated this ratio in a sample of 88 metal-poor halo stars, and the star with lowest value had $\log \epsilon$ (La/Eu)$=-0.01$; all three of our stars appear to have even smaller amounts of s-process material in their atmospheres. As s-process enrichment occurs on much longer timescales than r-process enrichment, this could be further evidence that the metal-poor bulge stars formed at earlier times than their halo counterparts.

\citet{2010ApJ...708.1398T} suggested that bulge metal-poor stars would show a wide scatter in r-process elemental abundances, even wider than that seen in halo stars at the same metallicity. The best r-process indicator element we have observed is Eu, and we do see a large scatter amongst our data points. Unfortunately, due to the Eu lines being quite weak in metal-poor stars, it was only observable in seven stars. No heavily r-process enhanced star has been uncovered. A greater number of metal-poor bulge stars would be needed to confirm the large [Eu/Fe] scatter.

\section{Conclusions}
The EMBLA survey is the first dedicated search for metal-poor stars in the Milky Way bulge. In this paper we have presented an abundance analysis of 10 stars observed with high-resolution spectroscopy using MIKE/Magellan in 2012; we have also reanalysed the four stars observed as part of Gaia-ESO, originally discussed in \citet{2014MNRAS.445.4241H}, to enable a homogeneous comparison. Ten of these have been confirmed spectroscopically as having metallicities of [Fe/H]$<-2.0$. Models of the formation of the Milky Way predict that the oldest stars should now be found in the Galaxy's most central regions (e.g. \citealt{2010MNRAS.401L...5S, 2010ApJ...708.1398T}), suggesting that the stars presented in this paper could be the oldest known.  

We have confirmed that the majority of the observed stars lie within the bulge region, given the distance uncertainties involved. Further, despite the lack of accurate proper motions of stars at these distances in the bulge, we have been able to show that the range of Galactocentric velocities shown by these stars is not dissimilar from other published bulge star samples. With the publication of upcoming data from OGLE, VVV, and ultimately Gaia, we will be able to confirm definitively whether these metal-poor bulge stars lie on tightly-bound orbits, as predicted by \citet{2010ApJ...708.1398T}, instead of merely being halo stars that are passing through the bulge region

The abundances of the stars do not deviate largely from those observed in the halo; however there are a few potentially crucial differences. All the stars have [C/Fe]$<0.7$, and this lack of carbon enhanced stars in the sample is unexpected. It can not be explained by internal mixing processes occurring due to the stars' evolved state, but perhaps could be due to a mix of selection effects (a study into the SkyMapper EMP survey and the lack of stars with [C/Fe]$<2.0$ will investigate this further), and the small sample size presented here.  

In \citet{2014MNRAS.445.4241H}, we discovered that the bulge stars may have a larger scatter in $\alpha$ abundances than expected, and this result is largely supported with our analysis here. The dispersion of Mg, Si and Ti are all approximately double that found in either more metal-rich bulge samples \citep{2013A&A...549A.147B}, or halo samples of similar metallicity \citep{2013ApJ...762...26Y}. The Ca abundances, however, do not show the same scatter. We also find that for three of the four elements, the [$\alpha$/Fe] values in these stars are slightly lower than for corresponding halo stars.

Wide scatter is seen in both r-  and s-process elements, but the ratios of Y and Sr to Ba do not support any signs of a previous generation of fast-rotating massive stars \citep{Cescutti:2013hl}. Furthermore, the ratios of La to Eu hint that there is less s-process enrichment in these stars than in halo metal-poor stars. Despite the small sample size, there are several stars with anomalous abundance results. One such star has [Sr/Ba]$=-0.78$, as well as the highest C abundance of the sample, [C/Fe]$=0.68$. Potentially this star could be in a binary system, where mass from a partner AGB star has been transferred. Other stars have anomalously high and low $\alpha$ abundances, for example, one star with [Mg/Fe]$=-0.18$ and [Ti/Fe]$=0.86$.

It is clear from these initial findings that the metal-poor stars in the bulge do have some chemical differences from those found in the halo, providing evidence for having formed at a different time in the history of the Universe. Crucially, in order to confirm that these stars are different from halo metal-poor stars, better kinematic data are required. If bulge-like orbits can be confirmed, these stars could give us a look at the earliest epochs in the life of the Milky Way.

\section*{Acknowledgments}
L.M.H. and M.A. have been supported by the Australian Research Council (grant FL110100012). A.R.C. acknowledges support from the European Union FP7 programme through ERC grant number 320360. DY is supported through an Australian Research Council Future Fellowship (FT140100554). Research on metal-poor stars with SkyMapper is supported through Australian Research Council Discovery Projects grants DP120101237 and DP150103294 (PI: Da Costa). This publication makes use of data products from the Two Micron All Sky Survey, which is a joint project of the University of Massachusetts and the Infrared Processing and Analysis Center/California Institute of Technology, funded by the National Aeronautics and Space Administration and the National Science Foundation.

\bibliography{references}
\bibliographystyle{mn2e2}

\bsp

\label{lastpage}
\end{document}